\documentclass[letterpaper]{vldb}

\usepackage{graphicx}
\usepackage{balance}
\usepackage{xspace}

\usepackage{times}
\usepackage{color}

\usepackage{url}
\usepackage{enumitem}
\usepackage{balance}

\usepackage{algorithm}
\usepackage{algorithmic}

\newcommand{\eml}{\texttt{ease.ml}\xspace}
\newcommand{\rr}{\textsc{RoundRobin}\xspace}
\newcommand{\hybrid}{\textsc{Hybrid}\xspace}
\newcommand{\greedy}{\textsc{Greedy}\xspace}

\newtheorem{theorem}{Theorem}

\newtheorem{lemma}[theorem]{Lemma}

\newcommand{\cN}{\mathcal{N}}

\newcommand{\E}{\mathbb{E}}

\newcommand{\bP}{\mathbb{P}} 

\DeclareMathOperator*{\argmax}{arg\,max}
\DeclareMathOperator{\SYN}{SYN}

\begin{document}


\title{Ease.ml: Towards Multi-tenant Resource Sharing for \\
Machine Learning Workloads}

\author{
\vspace{-1.1em}
~\\
Tian Li$^\dagger$~~~~~Jie Zhong$^\ddagger$~~~~~Ji Liu$^\ddagger$~~~~~Wentao Wu$^*$~~~~~Ce Zhang$^\dagger$\\
\small $^\dagger$ETH Zurich, Switzerland\\
\small litian@ethz.ch~~~~~ce.zhang@inf.ethz.ch\\
\small $^\ddagger$University of Rochester, USA\\
\small jie.zhong@rochester.edu~~~~~jliu@cs.rochester.edu\\
\small $^*$Microsoft Research, USA\\
\small wentao.wu@microsoft.com
}

\maketitle

\begin{abstract}
We present \eml, a \emph{declarative} machine learning service platform we built
to support more than ten research groups outside the computer science departments at ETH Zurich 
for their machine learning needs.
With \eml, a user defines the high-level 
schema of a machine learning application and submits
the task via a Web interface. The system automatically deals with the rest, such as model selection and data movement.
In this paper, we describe the \eml architecture and focus on
a novel technical problem introduced by \eml regarding resource allocation.
We ask, {\em as a ``service provider'' that manages a shared cluster of 
machines among all our users running machine learning workloads, what is the resource
allocation strategy that maximizes the global satisfaction of all
our users?}

Resource allocation is a critical yet subtle issue in this multi-tenant scenario, as we 
have to balance between efficiency and fairness.
We first formalize the problem that we call {\em multi-tenant model selection}, 
aiming for minimizing the total {\em regret} of all users
running automatic model selection tasks.
We then develop a novel algorithm that combines multi-armed bandits with Bayesian 
optimization and prove a regret bound 
under the multi-tenant setting.
Finally, we report our evaluation of \eml on synthetic data and on one service 
we are providing to our users, namely, 
image classification with deep neural networks.
Our experimental evaluation results show that our proposed solution can 
be up to 9.8$\times$ faster in achieving the same global quality
for all users as the 
two popular heuristics used by our users before \eml.

\end{abstract} 

\vspace{-1em}
\section{Introduction}\label{sec:introduction}

The past decade has witnessed the increasingly 
ubiquitous application of machine learning techniques 
in areas beyond computer sciences. 
One consequence of this wider application
is that we can no longer assume a computer science 
background on the part of our users. As a result, how to make
machine learning techniques more {\em accessible} and
{\em usable} to non-computer science users has become
a research topic that has attracted intensive interest
from the database community~\cite{Bailis2017,Binnig2016,Krishnan2017,Tamagnini2017,Varma2017}.

\paragraph*{Motivating Example} At ETH Zurich,
we as a research group provide ``data science
services'' to other research groups outside the computer science domain.
These users provide
us their data sets and the tasks they 
want to perform with machine learning. We host these data sets on our machines
for our users to run their machine learning
jobs.
Although we also have students serving as ``consultants''
to answer user questions, it is still our
users who run the machine learning systems
by themselves in most cases. As of today, our infrastructure
contains 24 TITAN X GPUs and 100 TB storage
shared by more than ten research groups
from areas that include astrophysics,
biology, social sciences, meteorology, 
and material science.
In this paper, we ask this question: {\em What is an
efficient and effective way to enable multiple users sharing 
the same computational infrastructure to run
machine learning tasks?}

\vspace{0.1em}
\noindent
{\bf Failed Experience 1.} Our first strategy was to provide
all users with \texttt{ssh} access to all our machines
and a shared Google Calendar for resource allocation.
However, even when everyone respects the resource allocation
protocol (which rarely happens), our users compete for 
resources fiercely. This decentralized management strategy fell into chaos and failed in less than two weeks. 

\vspace{0.1em}
\noindent
{\bf Failed Experience 2.} We then resorted to classic resource
managers and used \texttt{Slurm} for users to submit 
their jobs. 
Although this strategy isolates
users from competing for resources, it poses a
new problem for effective resource utilization.
In almost all our applications, the user needs to conduct 
a series of explorations of different machine learning
models.
However, not all the explorations conducted by
our users are necessary (either because the users lack
machine learning knowledge or the explorations are automatic 
scripts conducting exhaustive searches). For example,
one of our users used five GPUs for a whole week
trying different models to improve
a model that already had an accuracy of 0.99.
Another user 
continued to use his resources, trying deeper and 
deeper neural networks even though much simpler
networks already overfit on his data set.
If these resources were allocated to other users,
it would result in much better use of the computation time.

\vspace{0.1em}
Motivated by these experiences, we designed
\eml, a declarative machine learning service platform we built for our local collaborators at ETH Zurich
that employs an automatic resource manager for multi-tenant machine learning workloads.
Compared to existing multi-tenant 
resource managers~\cite{DavidShue2012,JonathanMace2015,Krebs2014},
\eml is aware of the underlying workload and is able to
integrate knowledge about machine learning
to guide the exploration process more effectively.
Compared to existing systems built for the
single-tenant case
such as Auto-WEKA~\cite{Kotthoff2017,Thornton2013}, 
Google Vizier~\cite{Golovin2017}, Spark TuPAQ~\cite{Sparks2015}, and Spearmint~\cite{Snoek2012}, \eml allows multiple users to share the same
infrastructure and then tries to balance and optimize their use of it for
the ``global satisfaction'' of all users.

\vspace{-0.5em}
\paragraph*{Challenges}

Automatic model selection is crucial to declarative machine learning services 
and has been studied intensively and extensively in the literature.
To enable automatic model selection for multiple users
competing for a shared pool of resources,
we started from two previous approaches:
(1) single-tenant model selection using 
\emph{multi-armed bandits}~\cite{Feurer2015,Golovin2017,Kotthoff2017,Snoek2012,Sparks2015};
and (2) {\em multi-task} Bayesian optimization
and Gaussian Process~\cite{Bardenet2013,Hutter2011,Swersky2013}.
When we tried to extend these methods to the particular multi-tenant scenario \eml was designed for,
we faced two challenges.

\noindent
{\bf (Optimization Objective)}
The optimization objective of single-tenant model selection is clear: Find the best model for the user as soon as 
possible (i.e., minimize the user's {\em regret} accumulated over time).
However, the optimization objective becomes messy once we turn to the multi-tenant case.
Previous work has proposed different objectives~\cite{Swersky2013}.
Unfortunately, none of these fit \eml's application scenario (see Section~\ref{sec:relatedwork}).
Therefore, our first challenge was to design an appropriate objective for \eml that captures the intuitive yet vague notion of ``global satisfaction'' for all users and then design 
an algorithm for this new objective.

\noindent
{\bf (Heterogeneous Costs and Performance)} Most
existing single-tenant model selection
 algorithms are aware of the execution cost of
  a given model. This is important as models with
  vastly different costs may have similar performance 
  on a particular data set. Therefore, any multi-tenant
  model selection algorithm that is useful in practice should also
  be aware of costs.
  Previous work on cost-aware model-selection resorts to various heuristics to integrate the cost~\cite{Snoek2012}.
  Moreover, theoretical analysis of many state-of-the-art algorithms is usually done in a ``cost-oblivious'' setting where costs are neglected.
  The question of how to integrate costs into model-selection algorithms while retaining the desirable theoretical properties remains open.
  Our second challenge was to develop cost-aware multi-tenant algorithms with theoretical guarantees.
  
\vspace{-0.5em}
\paragraph*{Summary of Technical Contributions}

With the above challenges in mind, we developed a novel framework for multi-tenant, cost-aware model selection and integrated it into \eml. We summarize our contributions as follows.

\noindent
{\bf C1. (System Architecture and Problem Formulation)}
Our first contribution is the architecture of \eml and the formulation of its core technical problem.
To the best of our knowledge, \eml is one of the first systems that provides declarative machine learning services.
In \eml, a user thinks about machine learning as an {\em arbitrary function approximator}.
To specify such an approximator, the user provides the
system with (1) the size of the input, (2) the size of the output,
and (3) pairs of examples that the function aims to approximate.
For example, if the user wants to train a classifier
for images into three classes, she would submit a job like
\vspace{-0.5em}
\[
\texttt{Input = [256, 256, 3]
        Output = [3]}.
\]
\eml then automatically matches all {\em consistent} machine learning models (e.g., AlexNet, ResNet-18, GoogLeNet, etc.) 
that can be used for this job and explores these models.
Whenever a new model finishes running, \eml returns the model to the user.
Despite the simplicity of this abstraction, more than 70\% of
our users' applications can be built in this way.

The core of the \eml architecture is an automatic scheduler that prioritizes the execution of different models for different users.
We further formalize the scheduling problem as what we call {\em multi-tenant model selection}.
Although similar problems have been explored in previous work~\cite{Swersky2013}, to the best of our knowledge we are the first to formulate the multi-tenant model selection problem in the context of a multi-tenant machine learning platform.
As a consequence, we have come up with a new optimization objective based on real user requirements in practice.

\noindent
{\bf C2. (Multi-tenant Model Selection Algorithms)} Our second contribution
is a novel algorithm for 
multi-tenant, cost-aware model selection.
Our proposed algorithm adopts a two-phase approach.
At each time step, it first determines the 
best model to run next for each user by
estimating the ``potential for quality improvement'' for each model.
It then picks the user with the highest potential in terms of the estimated quality improvement.
For the first ``model-picking'' phase, we developed a cost-aware variant of the 
standard GP-UCB algorithm~\cite{SrinivasKKS10} for selecting the best model of 
each user. For the second ``user-picking'' phase, we developed a simple but novel criterion to decide on the best user to be scheduled next.

We studied the theoretical properties of the proposed algorithm as well as a variant that replaces the criterion in the ``user-picking'' phase with a round-robin strategy.
In summary, we proved rigorous regret bounds for both algorithms: For the multi-tenant model selection problem 
with $n$ users (each user has $K$ candidate models to choose from), 
the total regret $R_T$ of all users is bounded by
\vspace{-0.25em}
\[
R_T < C n^{3/2}\sqrt{T \log(KT^2)\log(T/n)},
\]
where $T$ is the total execution time
and $C$ is a constant.
This is in line with the best-known bound for the standard GP-UCB algorithm.
It implies that both algorithms are \emph{regret-free}, i.e., $R_T/T \rightarrow 0$, which is a desired property for any practical algorithm.

We further analyzed the strength and weakness of both algorithms and designed a hybrid algorithm that explicitly considers
the varying accuracy of the underlying estimator employed by the ``user-picking'' criterion.
The hybrid algorithm observes the same regret bound but outperforms both of the original algorithms.

\noindent
{\bf C3. (Evaluation)} Our third contribution is an extensive
evaluation of \eml on synthetic data and on a real service that we are currently providing to our users: image classification 
with neural networks.
When a user submits a job with the schema
that maps a three-dimensional tensor to
a vector, \eml automatically matches the job with eight
different neural network architectures, including
NIN, GoogLeNet, ResNet-50, AlexNet,
BN-AlexNet, ResNet-18, VGG-16,
and SqueezeNet. We collect twenty 
data sets (users) and compare
\eml's scheduler to the popular heuristics 
used by our users (prior to the availability of \eml), as well as variants of our algorithm that leverage classic scheduling policies in the ``user-picking'' phase such as round robin or random scheduling.
We show that \eml is able to outperform
these workload-agnostic schedulers
by up to 9.8$\times$.
On synthetic data sets, we observe similar behaviors and speedups.

\vspace{-0.75em}
\paragraph*{Limitations and Future Work}
As of today, \eml has been running to support 
applications for our local collaborators~\cite{Schawinski2017}. 
As the number of users growing, we
expect that the current framework, both in a
theoretical and a practical perspective,
needs to be improved. We highlight these
limitations and promising future work in 
Section~\ref{sec:limitations}. The multi-tenant
model selection in \eml is motivated by
our experience in supporting our users, but 
the applicability goes beyond \eml --- we hope
this framework can also help other, much larger, service providers
of machine learning infrastructures such as 
Azure Machine Learning Studio
and Amazon Machine Learning
to reduce their operating
cost.

\vspace{-0.75em}
\paragraph*{Overview}
The rest of this paper is organized as follows.
We describe the system architecture of \eml in Section~\ref{sec:architecture}.
As a preliminary, we present the single-tenant model selection problem and a cost-aware variant of the standard GP-UCB algorithm in Section~\ref{sec:singletenant}.
We then formalize the multi-tenant model selection problem and present our solution and theoretical analysis in Section~\ref{sec:multitenant}.
We report experimental evaluation results in Section~\ref{sec:experiments}, summarize related work in Section~\ref{sec:relatedwork}, and conclude the paper in Section~\ref{sec:conclusion}.

\begin{figure*}
\centering
\includegraphics[width=0.8\textwidth]{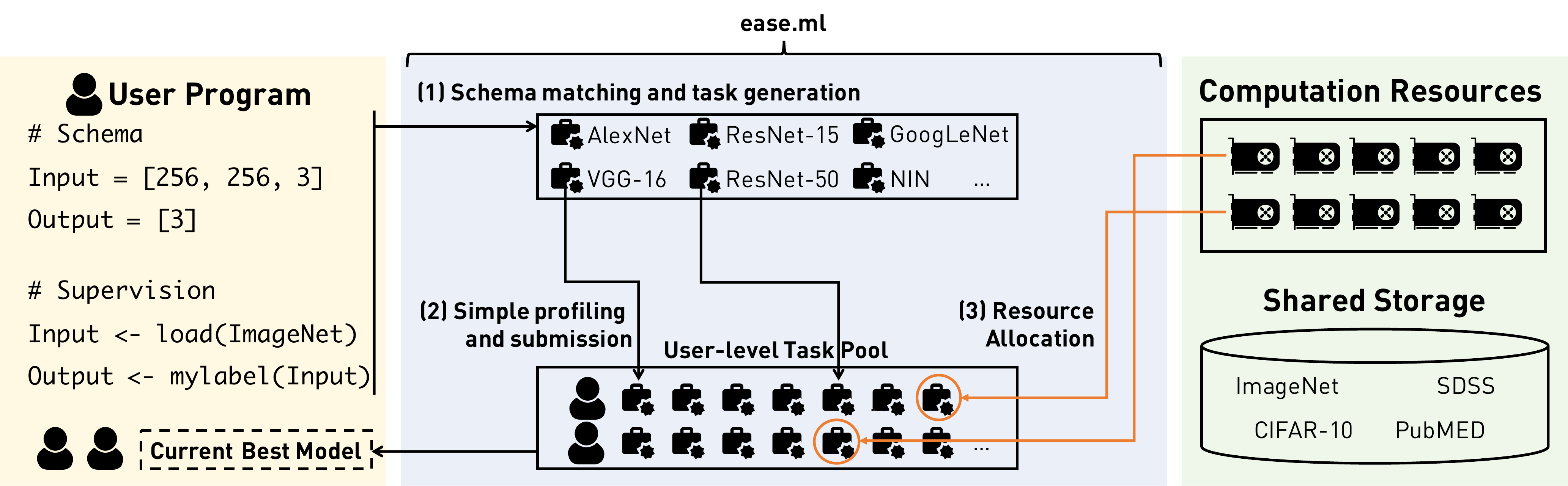}
\vspace{-1em}
\caption{System architecture of \texttt{ease.ml}.}
\label{fig:architecture}
\vspace{-1em}
\end{figure*}

\begin{figure}[t]
\centering
\includegraphics[width=0.5\textwidth]{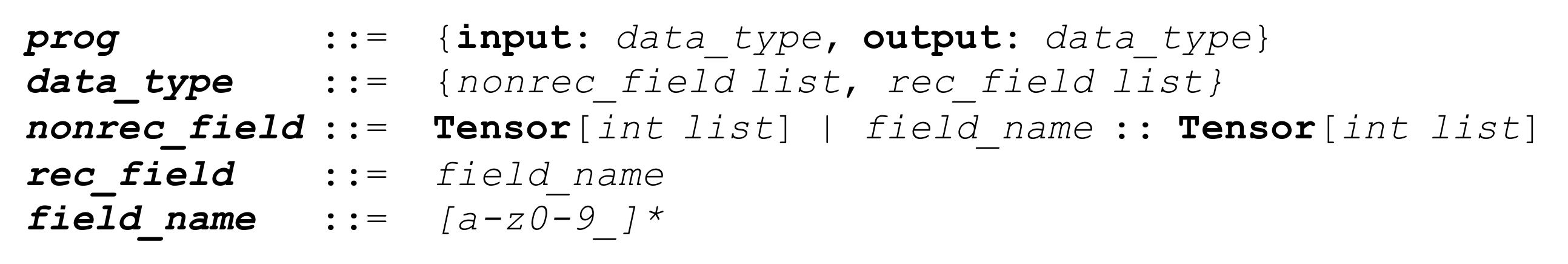}
\vspace{-2em}
\caption{Formal syntax of an \eml program}
\label{fig:syntax}
\vspace{-1em}
\end{figure}

\newpage
\vspace{-1em}
\section{System Architecture} \label{sec:architecture}

The design goal of \eml is twofold: (1) provide an abstraction
to enable more effective model exploration for our users, and 
(2) manage the shared infrastructure to enable more efficient
resource utilization during the exploration process 
for {\em all} instead of {\em one} of our users.
A similar, but vague, objective for \eml was published as a short vision paper~\cite{Zhang2017a}. This paper tackles the first
concrete technical problem we faced in realizing this
vision.

\eml provides a simple high-level declarative language.
In \eml, the user thinks about machine learning as an {\em arbitrary function approximator}.
To specify such an approximator, the user provides the
system with (1) the size of the input, (2) the size of the output,
and (3) pairs of examples that the function aims to approximate.
Figure~\ref{fig:architecture} illustrates the system architecture of \eml.

\vspace{-0.5em}
\subsection{Components and Implementations}

We walk through each component of \eml and 
describe design decisions
motivated by our observation of our users. 

\vspace{-1em}
\paragraph*{Input Program} The input
of \eml is program written in 
a simple domain-specific language (DSL). Figure~\ref{fig:syntax}
shows the formal syntax of the DSL. To specify
a {\em machine learning task}, the user program (\textbf{\texttt{\em prog}})
contains the information about the {\em shape}
and {\em structure} of an input object (e.g., images, time series)
and an output object (e.g., class vector, images, time series).

The design goal of input and output
objects is to provide enough flexibility to support most
workloads that we observed from our users. In the current design,
each object (\textbf{\texttt{\em data\_type}}) contains two parts:
(1) the ``recursive'' component (\textbf{\texttt{\em rec\_field}})
and (2) the ``nonrecursive'' component (\textbf{\texttt{\em nonrec\_field}}).
The recursive component contains a list of named 
fields of the type of the same object, and the nonrecursive
component contains a list of constant-sized tensors.
This combination of recursive and nonrecursive
components allows \eml to model a range of the workloads that our users need, including image, time series, and trees.

\begin{figure}[t]
\centering
\includegraphics[width=0.5\textwidth]{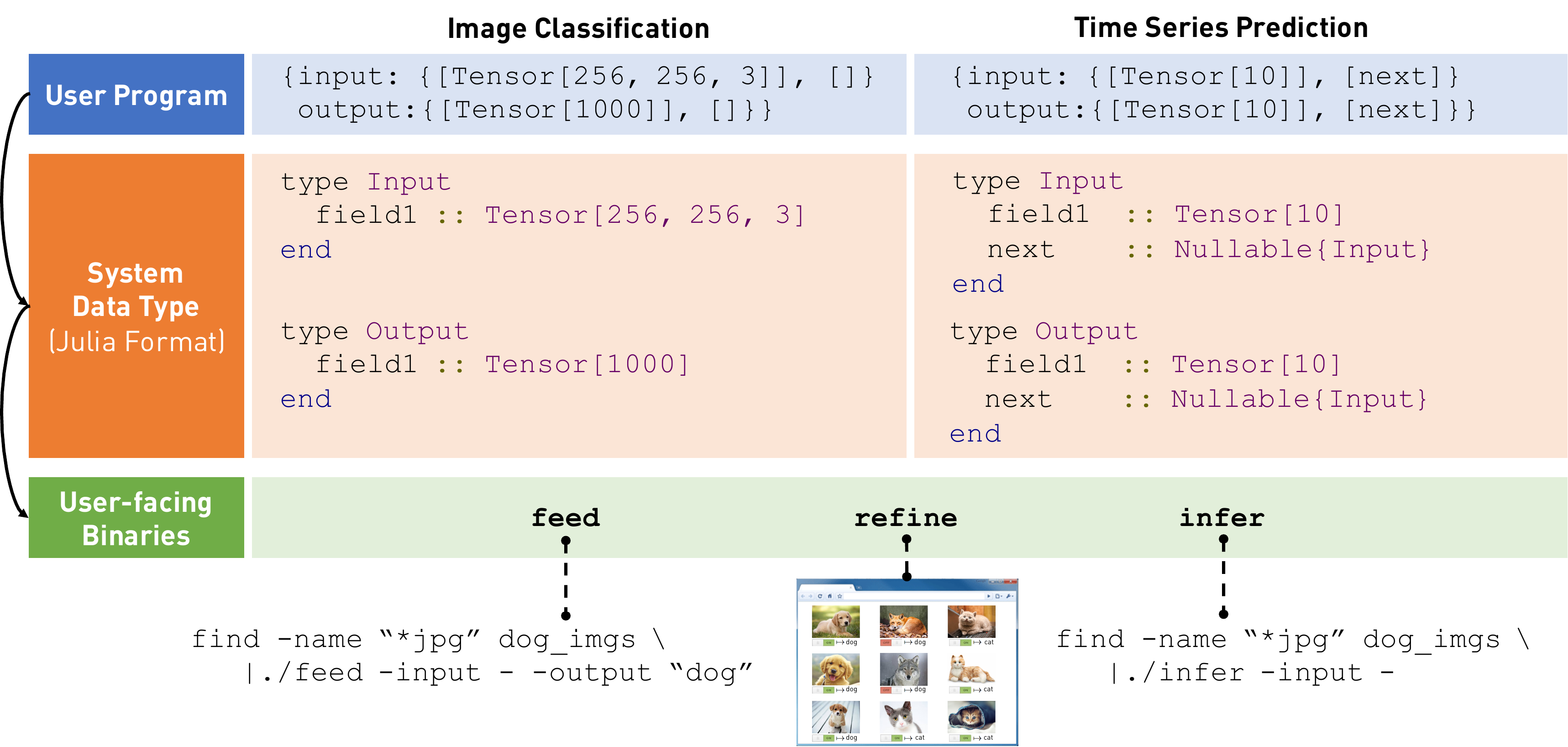}
\vspace{-2em}
\caption{System walkthrough}
\vspace{-2em}
\label{fig:walkthrough}
\end{figure}

\vspace{0.5em}
\noindent
{\bf (Example)} Figure~\ref{fig:walkthrough} shows
two example user programs for (1) image classification and (2) time series
prediction. For image classification, each input
object is a tensor of the size 256$\times$256$\times$3 and
each output object is a tensor of the size 1,000 corresponding
to 1,000 classes. Here the input
and output objects contain only the nonrecursive component.
For the time series prediction, each object 
contains a 1-D tensor and 
a ``pointer'' to another object of the same type. This
forms a time series.

\vspace{-2em}
\paragraph*{Code Generation and User Interaction}

Given an input program, \eml conducts code generation
to generate binaries that the user directly interacts with.
Figure~\ref{fig:walkthrough} illustrates the process.

In the first step of code generation, \eml translates
the user programs into {\em system-data types}, data types
that the rest of the system is able to understand. 
Figure~\ref{fig:walkthrough} shows the system-data types
generated in Julia. The translation process is based
on very simple operational semantics, and we thus omit the details
here. One inherent assumption during the translation
is that there is no reuse of objects, i.e., we can 
only generate types corresponding to DAG without loop
(e.g., singleton, chains, and trees). 

Given the system-data types generated by the translation
procedure, \eml then generates three binaries and
a Python library.
The Python library shares the same functionality as
the binaries but can be used in a programmable way.
Figure~\ref{fig:walkthrough} shows the three binaries,
and Figure~\ref{fig:architecture} illustrates one
example in Python. The binaries and the Python library
contain a unique identifier and an IP address mapped to the
\eml server. All operations the users conduct with
the generated binaries and Python library will be 
sent to the server, which hosts the shared
storage and the pool of computation resources.
There are three basic operations in \eml.

\noindent
{\textbf{1.~\texttt{feed}}.} The \texttt{feed} operator takes
as input a set of input/output pairs. \eml provides
a default loader for some popular Tensor types (e.g., loads
JPEG images into Tensor[A,B,3]). To associate an output
object and an input object, the user can simply pipe a pair of
objects into the binary, as shown in Figure~\ref{fig:walkthrough} or
write a {\em labeling function} that maps an input
object into an output object as shown in Figure~\ref{fig:architecture}.
Every time the user invokes the \texttt{feed} operator,
all data will be sent and stored in the centralized
\eml server.

\noindent
{\textbf{2.~\texttt{refine}}.} The \texttt{refine} operator
shows all input/output pairs that the user ever \texttt{feed}'ed
into the system and allows the user to ``turn on''
and ``turn off'' each example. This is especially useful when
the users want to conduct data cleaning on the training set
to get rid of noisy labels introduced by weak or distant supervision.

\noindent
{\textbf{3.~\texttt{infer}}.} The \texttt{infer} operator
takes as input an input object and outputs an output object
using the {\em best model learned so far}.

\begin{figure}[t]
\centering
\includegraphics[width=0.5\textwidth]{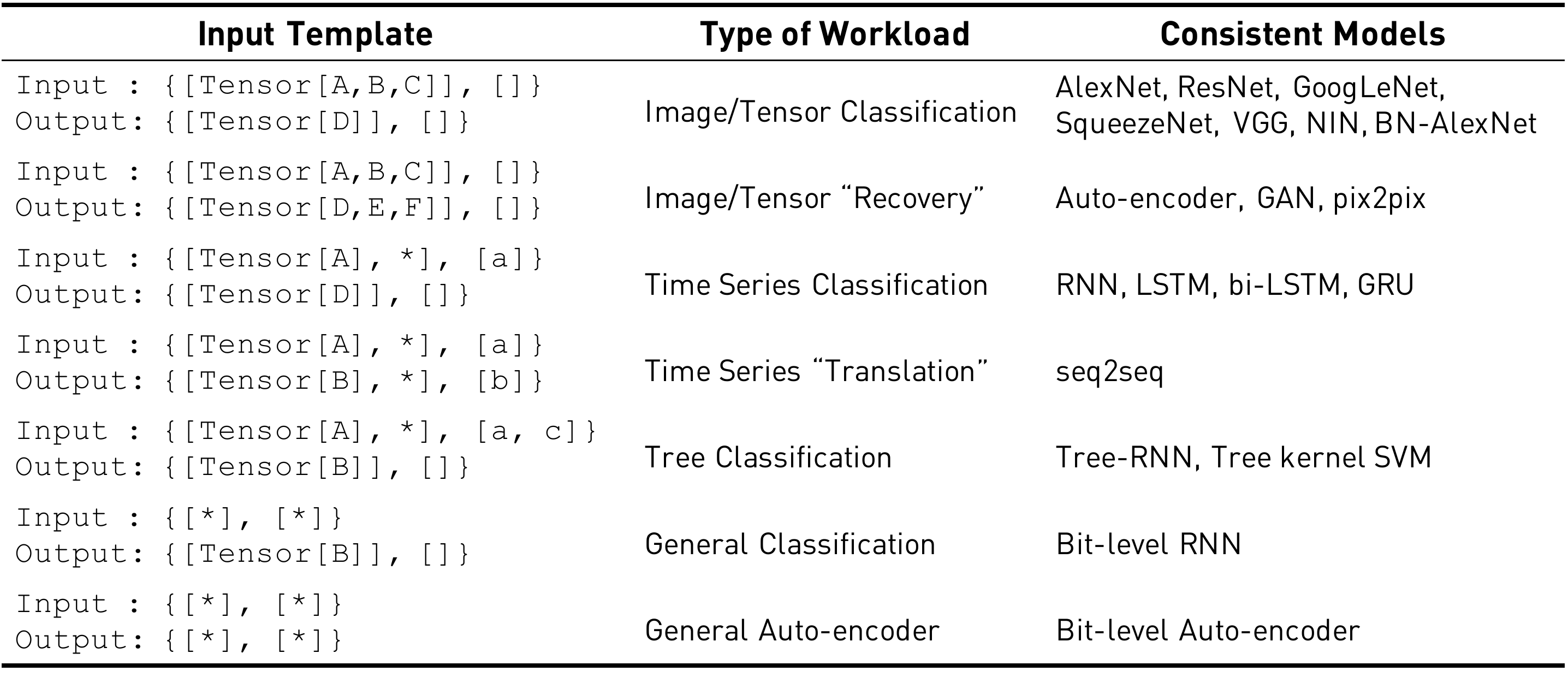}
\vspace{-2em}
\caption{Templates for candidate model generations. \texttt{A},
\texttt{B}, \texttt{C}, \texttt{D}, \texttt{E}, and \texttt{F}
are natural number constants; \texttt{a}, \texttt{b}, and \texttt{c} are of the type \texttt{\em field\_name}. \texttt{*} represents
matching for arbitrary ``tail'' of an array. Matching order
goes from top to bottom.}
\label{fig:templates}
\vspace{-1em}
\end{figure}

\paragraph*{Automatic Model Exploration}

The above interface provides the user with a high-level abstraction
in which the user only has a ``view'' of the best available model
instead of {\em what} model the system trains and 
{\em when} and {\em where} a model is trained. To enable 
this interface, automatic model selection plays a
central role.

\vspace{0.5em}
\noindent
{\bf (Candidate Model Generation: Template Matching)} 
The first step of automatic
model exploration is to generate a set of 
candidate models given a user program. The current
version of \eml uses a template-matching approach.
Figure~\ref{fig:templates} shows the current set of
templates and the corresponding candidate models. 
The matching happens from the top to the bottom 
(from the more specific template to the more general template).

\begin{figure}[t]
\centering
\includegraphics[width=0.3\textwidth]{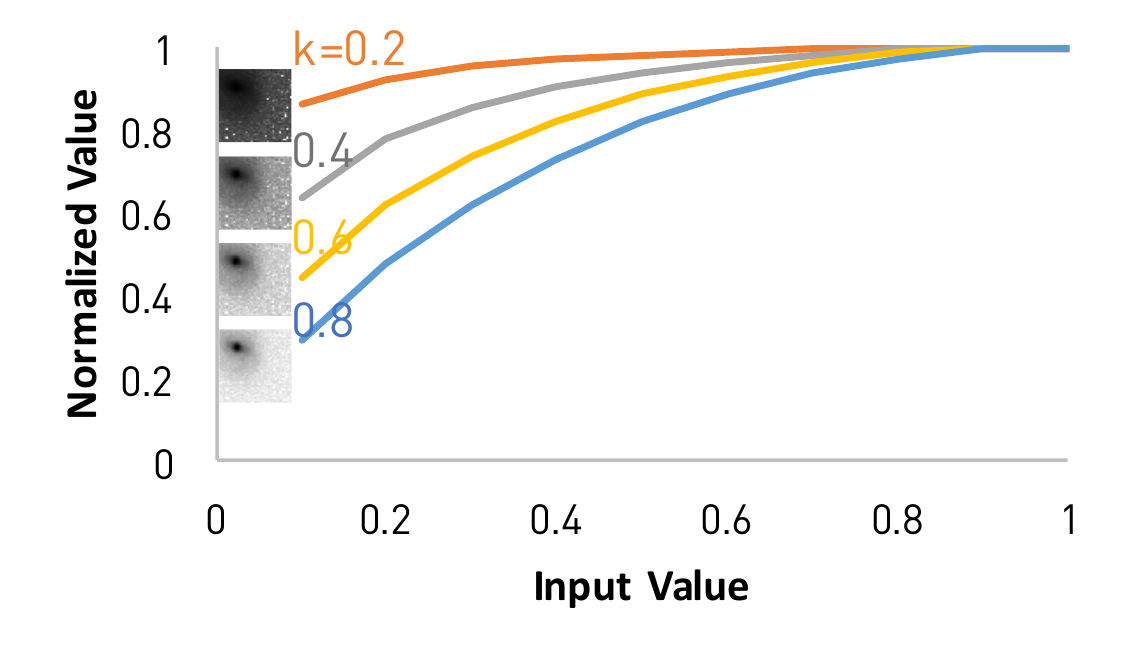}
\vspace{-2em}
\caption{Automatic normalization in \eml. Normalization function
is from $f_k(x) = -x^{2k} + x^{k}$ with different 
values of $k$. Each $k$ generates one additional candidate
model. Snapshots of a galaxy illustrate the impact of
normalization. Images are from the astrophysics 
application motivated this feature.}
 \label{fig:normalization}
 \vskip -2ex
\end{figure}

\vspace{0.5em}
\noindent
{\bf (Candidate Model Generation: Automatic Normalization)} 
In addition to template matching, another source of candidate models 
comes from \eml's automatic normalization feature.
For image-shaped templates (e.g., Tensor[A,B,3]),
most of the consistent deep-learning models are designed for
image data. However, much of the data from our users, although
they have shapes like an image, have a much larger dynamic
range than an image. For one astrophysics application~\cite{Schawinski2017}
and one proteomics application, the dynamic range could
vary by more than ten orders of magnitude. In this case,
simply treating the input as an image results in unusable 
quality. \eml provides an automatic input
normalization feature by normalizing the input
with a family of functions as shown in Figure~\ref{fig:normalization}.
Each normalization function in this family, together with 
a consistent model, generates one  
candidate model.

\vspace{0.5em}
\noindent
{\bf (Automatic Model Selection)} 
Given a set of candidate models, \eml decides on an order of
execution. The current execution strategy of \eml is to 
use {\em all} its GPUs to train a single 
model (see Section~\ref{sec:limitations}
for a discussion about this design decision).
In the near 
future, \eml will need to allow a more flexible resource 
allocation strategy to support a resource pool 
with hundreds of GPUs. Because there are different
users using \eml at the same time, \eml also needs
to decide which user to serve at the current time.
This problem motivated the core technical problem of this paper,
which we describe in detail in Sections~\ref{sec:singletenant} and~\ref{sec:multitenant}.

\vspace{-1em}
\paragraph*{Discussion: Hyperparameter Tuning}
We highlight one design decision in \eml
that is not optimal. \eml also conducts automatic
hyperparameter tuning but treats it as part of the
training procedure of a model. For example, for the model-selection
subsystem, when it decides to train a given model,
it will always train with
automatic hyperparameter tuning. A more optimal design 
would be to fuse the hyperparameter tuning subsystem
with the model-selection subsystem to better
utilize available resources to maximize the
{\em global satisfaction of all users}.

\section{Single-tenant Model Selection} \label{sec:singletenant}

We now focus on the model-selection subsystem in \eml. In this section,
we discuss the single-tenant case in which a single user needs to
choose from $K$ different models. We first present background on
the classic, cost-oblivious GP-UCB algorithm and then present a simple
twist for a cost-aware GP-UCB algorithm. Despite the simplicity of
this twist, the resulting cost-aware algorithm forms a basic building
block for the multi-tenant setting.

\vspace{-1em}
\paragraph*{Problem Formulation} We treat
model selection as a multi-armed bandit problem.
Each algorithm corresponds to an \emph{arm} of 
the bandit, whereas the observed evaluation result of 
the algorithm corresponds to the \emph{reward} of playing the conceivable arm.
A common optimization criterion is to minimize the cumulative
\emph{regret}.

Formally, let there be $K$ arms and $[K]$ be the set of
all arms. Let $x_{a_t, t}$ be the reward of playing the arm $a_t\in[K]$
at time $t$, and suppose that $x_{a_t, t}$ follows a distribution with the mean $\mu_{a_t}$. 
Let $\mu_*=\max_{k\in[K]}\mathbb{E}(\mu_k)$ be the ``best'' solution in the expectation sense that is unknown to the algorithm.
The {\em instantaneous} regret at time $t$ if we
play the arm $a_t$ is $r_t=\mu_*-\E(x_{a_t, t})$ or equivalently $r_t = \mu_*-\mu_{a_t}$.
The {\em cumulative} regret up to time $T$ is defined as
\[
R_{T}=\sum\nolimits_{t=1}^{T}r_t = \sum\nolimits_{t=1}^{T}\Big(\mu_*-\E(x_{a_t, t})\Big).
\]
For different strategies of choosing different arms at
each time $t$, they incur different cumulative regret.
One desired property of the bandit problem is 
that asymptotically there is no cumulative regret, i.e.,
$\lim\nolimits_{T\to\infty}R_T/T=0.$ The performance
of a given strategy can be measured as the {\em convergence rate}
of the cumulative regret.


\vspace{-1em}
\paragraph*{Discussion: Relation to Model Selection} 
The randomness commonly exists in  machine learning algorithms. 
In our model-selection setting, we treat each arm as
a model and the {\em random} reward $x_{a_t, t}$ is the {\em accuracy} of
the model $a_t$ at time $t$. 
The best model that the user
has at time $t$ has the quality $x_t = \max_t x_{a_t, t}$. We
define a slightly different version of {\em regret} that is
directly associated with the user experience in \eml:
\[
R_{T}' = \sum\nolimits_{t=1}^{T} \left(\mu_* -  \mathbb{E}(\max_t x_{a_t, t})\right).
\]
Intuitively, this means that the user experience of a single
user in \eml relies not on the quality of the model the system
gets at time $t$ but on the best model so far up to time $t$ because
it is the ``best model so far'' that \eml is going to use 
to serve the \texttt{infer} operator. To
capture the relation between the ``\eml regret'' $R_T'$ and 
the classic cumulative regret $R_T$, observe that
\[
R_T' \leq R_T~~~~~\forall t, a_1,...,a_t.
\]
Because we are only interested in
the upper bound of $R_T'$, for the rest of this paper,
we will always try to find the upper bound for 
the standard cumulative regret $R_T$ instead of 
$R_T'$ directly.

\begin{algorithm}[t!]    
\scriptsize
\caption{Single-tenant, cost-oblivious GP-UCB}          
\label{alg:gp-ucb}                           
\begin{algorithmic}[1]                    
   \REQUIRE GP prior $\mu_0$, $\sigma$, $\Sigma$, and $\delta\in (0,1)$
   \ENSURE Return the best algorithm among all algorithms in $a_{[1:T]}$
   \STATE Initialize $\sigma_0 = \text{diag}(\Sigma)$
   \FOR {$t=1,2,\cdots, T$}
   \STATE $\beta_t \leftarrow \log (Kt^2/\delta)$\\
   \STATE $a_t \leftarrow \argmax_{k\in [K]}\mu_{t-1}(k) + \sqrt{\beta_t}\sigma_{t-1}(k)$\\
   \STATE Observe $y_t$ by playing bandit $a_t$.\\
   \STATE $\mu_t(k) = \Sigma_t(k)^\top (\Sigma_t + \sigma^2 I)^{-1} y_{[1:t]}$\\
   \STATE $\sigma_t^2(k) = \Sigma(k, k) - \Sigma_t(k)^\top(\Sigma_t + \sigma^2 I)^{-1} \Sigma_t(k)$
   \ENDFOR
\end{algorithmic}
\vspace{-0.5em}
\hrulefill\\
\vspace{0.5em}
\textbf{Notation}:
\begin{itemize}[noitemsep,topsep=0pt,parsep=0pt,partopsep=0pt]
\item $[K] = \{1, 2, \cdots, K\}$.
\item $y_{[1:t]} = \{y_1, \cdots, y_t\}$.
\item $\Sigma(i,j) = \Sigma_{i,j}$, for $i,j\in [K]$.
\item $\Sigma_t(k) = [\Sigma(a_1, k), \cdots \Sigma(a_t,k)]^\top, k\in [K]$.
\item $\Sigma_t = [\Sigma(i, j)]_{i,j\in a_{[1:t]}}$.
\end{itemize}
\end{algorithm}

\begin{figure}
\centering
\includegraphics[width=0.4\textwidth]{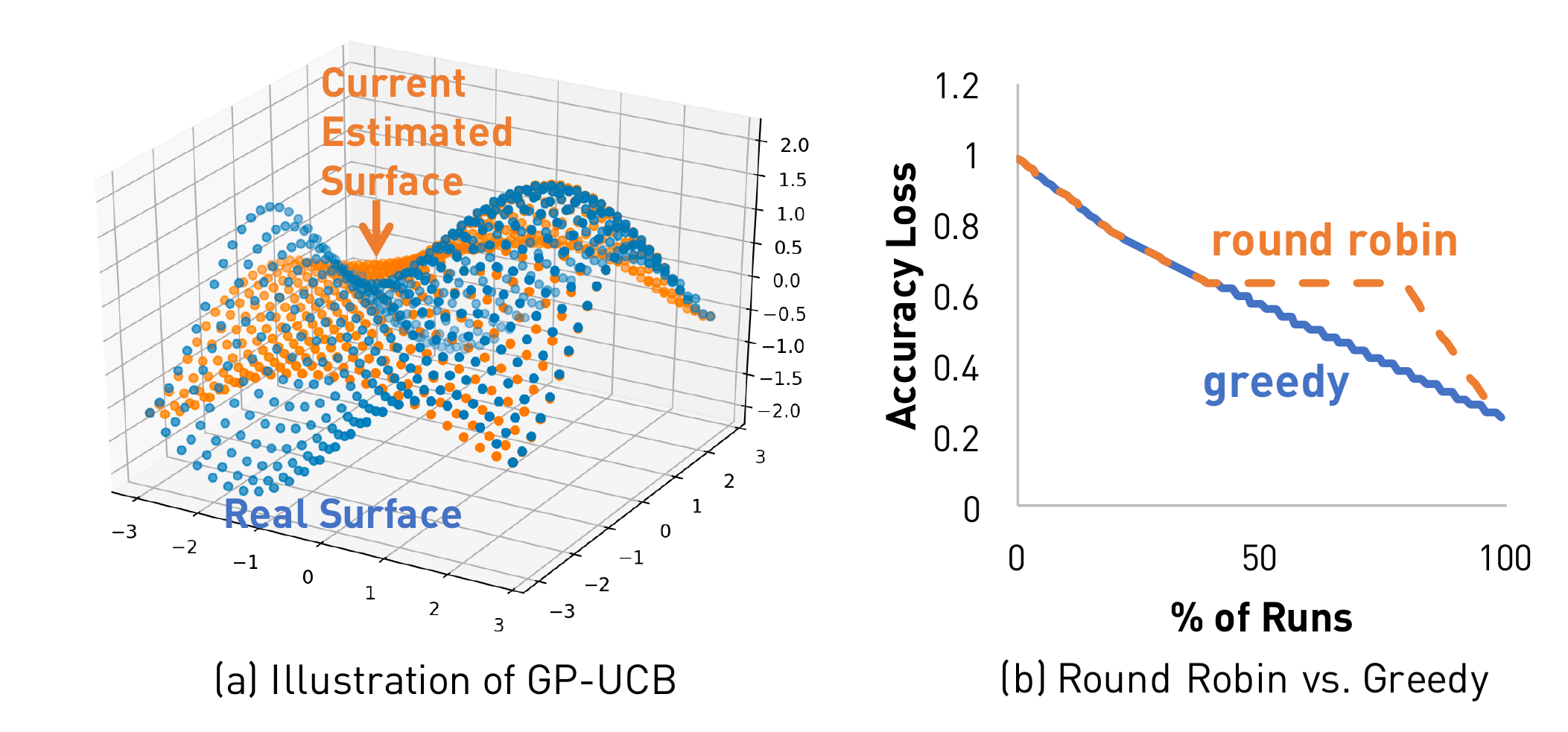}
\vspace{-1em}
\caption{Illustration of (1) Gaussian process and (2) the difference between \textsc{RoundRobin}
and \textsc{Greedy}.}
\label{fig:gpucb-intuition}
\vskip -2ex
\end{figure}

\subsection{Cost-Oblivious GP-UCB}

We start from the standard GP-UCB algorithm~\cite{DBLP:journals/ftml/BubeckC12} for a single-tenant bandit problem.
The key idea of GP-UCB is to combine the Gaussian Process (GP),
which models the belief at time $t$ on the reward of arms $\{x_{k,{t+1}}\}$
at the next timestamp, with the upper confidence bound (UCB) 
heuristic that chooses the next arm to play given the current belief.
Algorithm~\ref{alg:gp-ucb} presents the algorithm.

\vspace{-0.75em}
\paragraph*{Gaussian Process} The Guassian Process component model is
the reward of all arms at the next timestamp $t$ as a random draw from a 
Gaussian distribution
$\mathcal{N}(\mu_t, \tilde{\Sigma})$ where $\mu_t \in \mathbb{R}^K$ is the
mean vector and $\tilde{\Sigma} \in \mathbb{R}^{K \times K}$ is the covariance
matrix. Therefore, the marginal probability of each arm is also a Guassian:
In Algorithm~\ref{alg:gp-ucb} (lines 4, 6, and 7), the arm $k$ has distribution
$\mathcal{N}(\mu_t(k), \sigma_t^2(k))$. The belief of the joint distribution
keeps changing during the execution when more observations are made
available. Lines 6 and 7 of Algorithm~\ref{alg:gp-ucb} are used to
update this belief given a new observation $y_t$ (line 5).
Figure~\ref{fig:gpucb-intuition}(a) illustrates the Gaussian Process, in which
the blue surface is the real underlying function and the orange
surface is the current mean vector after multiple observations.

\vspace{-0.75em}
\paragraph*{Upper Confidence Bound (UCB)} The UCB rule
chooses the next arm to play with the following heuristic. At time $t-1$,
choose the arm with the largest $\mu_k + \theta\cdot\sigma_k$ where 
$\mu_k$ and $\sigma_k^2$ are the current mean and variance of the 
reward distribution of the arm $k$ (line 5 in Algorithm~\ref{alg:gp-ucb}). 
This is the upper bound of the $\theta$-confidence interval.
Despite its simplicity, it exhibits a trade-off between \emph{exploration} and \emph{exploitation}.
Intuitively, UCB favors arms with high reward (for exploitation) \emph{and} high uncertainty (for exploration), which implies high risk but also high opportunity for gaining a better reward.
The choice of $\theta$ has an impact on the convergence rate.

\vspace{-0.75em}
\paragraph*{Theoretical Analysis}

The well-known cumulative regret of the UCB algorithm~\cite{DBLP:journals/ftml/BubeckC12} is $R_T \leq C\cdot K\log T$,
where $C$ is some constant depending on the arm distribution. It is known to be one of the best upper bounds for the multi-bandit problem. This bound has a minor dependence on $T$ but depends seriously on the distribution of all arms and the number of arms. This is mainly because the UCB algorithm does not consider dependence between arms. In order to make $R_T/T$ converge to zero, we have to try at least $K$ times. So the UCB algorithm must play all arms once or twice in the initial step. To utilize the dependence between arms, previous work has extended the UCB algorithm with the Gaussian Process, where the dependence between arms can be measured by a \emph{kernel} matrix. The GP-UCB algorithm can achieve the regret
\[
R_T \leq C\cdot \sqrt{T\log(KT^2)\log T},
\]
where $C$ is some constant that does not depend on the distribution of arms. We can see that this bound only has minor dependence on the number of arms but has a greater dependence on $T$. GP-UCB does not have to pull all arms once to initialize the algorithm. It can achieve a satisfactory average regret before all arms get pulled. GP-UCB is suitable when the variance of each arm is small.

\vspace{-0.5em}
\subsection{Cost-aware Single-tenant GP-UCB}

The previous algorithm does not consider the {\em cost} of playing an arm.
In the context of model selection, this could correspond to the execution
time of a given model. 
As we will see, being aware of cost becomes even more important in the
multi-tenant setting as it is one of the criteria used to balance between
multiple users.

In this section, we extend the standard GP-UCB algorithm to a cost-aware version
with a very simple twist. We then analyze the theoretical property of
this cost-aware algorithm. We note here that, while the theoretical
analysis is relatively simple and thus we do not claim a technical contribution for
it, we did not see a similar algorithm or analysis in the literature.

\vspace{-0.75em}
\paragraph*{A Simple Twist} Based on Algorithm~\ref{alg:gp-ucb}, the idea is simple. 
We just replace Line 4 in the following (difference in red font)
\[
a_t \leftarrow \argmax_{k\in [K]}\mu_{t-1}(k) + \sqrt{\beta_t \textcolor{red}{/ c_k}}\sigma_{t-1}(k),
\]
where $c_k$ is arm $k$'s cost.
This introduces a trade-off between 
cost and confidence: Everything being equal, 
the slower models (larger $c_k$) have lower priority. However,
if it has very large potential reward (larger $\sigma_{t-1}(k)$),
even an expensive arm is worth a bet.

\vspace{-0.75em}
\paragraph*{Theoretical Analysis}
The following theorem bounds the cumulative regret
for our simple cost-aware GP-UCB extension.
We first define the cost-aware cumulative regret
as $\tilde{R}_T=\sum\nolimits_{t=1}^{T} c_{a_t} r_t$.

\vspace{-0.75em}
\begin{theorem}
  \label{thm:cost}
  If $c^\ast = \max_{k\in [K]}\{c_k\}$ and
  in Algorithm~\ref{alg:gp-ucb}
  $\beta_t = 2c^\ast \log\left[\frac{\pi^2 K t^2}{6\delta}\right]$, with probability at least $1-\delta$,
  the cumulative regret of single-tenant, cost-aware GP-UCB is bounded above by
  \[
  \tilde{R}_T < \sqrt{T\cdot I(T)}
  \]
where $
I(T) = \frac{4c^\ast \beta_T}{\log(1 + \sigma^{-2})}\sum_{t=1}^T \log(1+\sigma^{-2}\sigma^2_{t-1}(a_t))$.
Moreover, we have the bound for the instantaneous regret:
\[
    \bP\left(\min_{t\in [T]}r_t \le \sqrt{\frac{\tilde{I}(T)}{\sum_{t=1}^T
        c_{a_t}}}\right) \ge 1-\delta,
\]
where $\tilde{I}(T) = I(T)/c^\ast$.
\end{theorem}

\vspace{-0.5em}
Note that $I(T)$ is proportional to the information gain, and the order is at most $\log T$ (see Theorem 5 in \cite{SrinivasKKS10} for details). Therefore, this theorem is in line with the classical GP-UCB result and yields the desired \emph{regret-free} property that $R_T/T \to 0$ as $T\to \infty$.

In the same spirit, this theorem also suggests that the regret at iteration $t$ (the minimal regret up to iteration $t$) converges to $0$ with respect to the running time $\sum_{t=1}^T c_{a_t}$ with high probability.

\vspace{-1em}
\section{Multi-tenant Model Selection}\label{sec:multitenant}

We now present the multi-tenant model-selection
algorithm in \eml. This section is organized as follows.

\noindent
$\bullet$ We formulate the problem 
by extending the classic single-tenant regret
into a multi-tenant, cost-aware form. We then 
discuss the difference compared to
one very similar formulation~\cite{Swersky2013}.

\noindent
$\bullet$
We start from a very simple strategy that we call
\rr. \rr schedules
each user in a round-robin way while each user
uses their own single-tenant GP-UCB
during their allocated time slices. We proved 
a regret bound for this simple strategy.

\noindent
$\bullet$
We then present an improved algorithm that we
call \greedy. \greedy schedules
each user by maximizing their potential contribution
to the global objective. This is a novel algorithm.
We also prove a regret bound for this strategy.

\noindent
$\bullet$
The theoretical bound and empirical experiments
show a trade-off between \rr and
\greedy. Last, we present a \hybrid 
strategy that balances 
\rr and \greedy. \hybrid is the 
default multi-tenant model-selection algorithm
in \eml.

\vspace{-0.5em}
\subsection{Problem Formulation}

\begin{figure}[t]
\centering
\includegraphics[width=0.75\columnwidth]{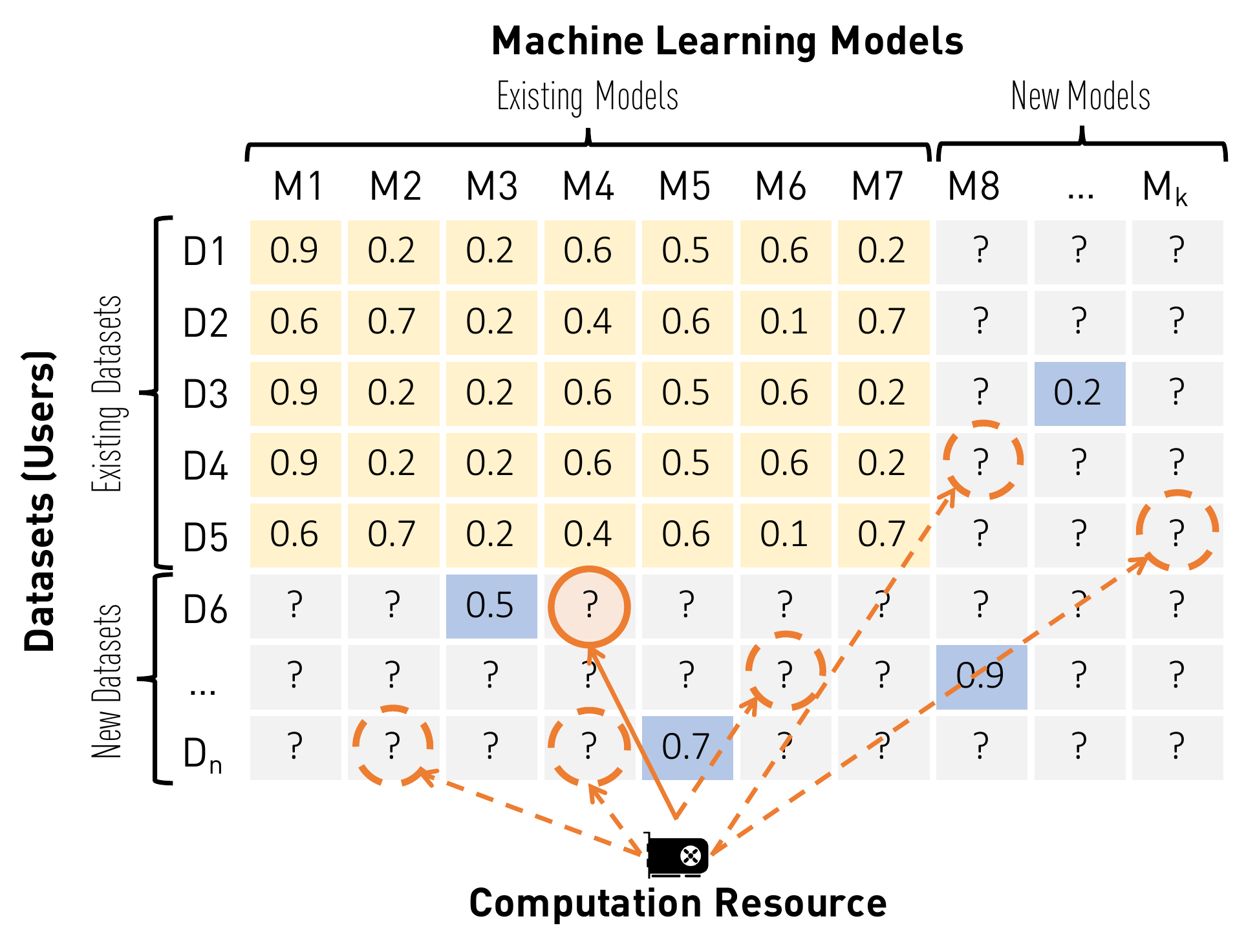}
\vspace{-2em}
\caption{A canonical view of the multi-tenant, cost-aware model-selection problem we studied in this paper.}
\label{fig:problem-formulation}
\vskip -4ex
\end{figure}

In the multi-tenant setting, \eml aims
to serve $n$ different users instead of a single user.
Without loss of generality, assume that
each user $i\in[n]$ has her own machine-learning 
task represented by a different data set
and that each user $i$ can choose
$K^{i}$ machine-learning models.
All users share the same infrastructure,
and {\em at a single time point only
a single user can be served}.
Figure~\ref{fig:problem-formulation} illustrates 
a canonical view of this problem.

\vspace{-0.5em}
\paragraph*{Multi-tenant Regrets} 
We extend the definitions of instantaneous and cumulative 
regrets for the multi-tenant setting. The key difference
compared to the single-tenant setting is that,
at round $t$, there are users who are not
served. In this case, how should we define
the regret for these unscheduled users?

The intuition is that these unscheduled users should
also incur a penalty --- because these users are not being served,
they do not have a chance to
get a better model. Instead, they need to stick
with the same model as before. 
Before we describe our extension for the
multi-tenant regret, we introduce notation
as follows. At round $t$,
\begin{enumerate}[noitemsep,topsep=0pt,parsep=0pt,partopsep=0pt]
\item $I_t$: The user that the system chooses to
serve.
\item $a_{t}^{I_t}$: The arm played by the chosen user $I_t$ at round $t$.
\item $t^i$: The last round a user $i$ gets served, i.e., $t^i = \argmax\{t': i=I_{t'}, 1\le t'\le t\}$.
\item $a_{t^i}^i$: The arm played by a user $i$ at the last round when she was served.
\item $c^i_k$: the cost of a tenant $i$ choosing a model $k\in [K^i]$.
\item $C_t :=c^{I_t}_{a_{t}^{I_t}}$: the cost of the algorithm chosen at round $t$.
\item $\mu_*^i$: the best possible quality that a user $i$ can get.
\item $X_t^i := x_{a_{t^i}^i, t^i}$: the rewards user $i$ gets at time $t^i$.
\end{enumerate}

We define the cumulative,
multi-tenant, cost-aware, regret as
\[
R_T = \sum_{t=1}^T C_t \left (\sum_{i=1}^n r^i_{t^i} \right),
\]
where $r^i_{t^i} = \mu_*^i - \mathbb{E}(X_t^i)$
is the regret of a user $i$ for continuing using the model chosen at the last time she
got served. 

\paragraph*{Ease.ml Regret} Similar to
the single-tenant case, we can define
a variant of the cumulative 
regret $R_t$ that directly relates to
\eml's design of always returning the
{\em best model so far}:
\[
R_T' = \sum_{t=1}^T
C_t
\left(
\sum_{i=1}^n 
\left(
\mu_*^i - \mathbb{E}(\max_t X_t^i)
\right)
\right) < R_T.
\]

\vspace{-1em}
\paragraph*{The Problem of ``First Come First Served''}
One of the most straightforward ideas of
serving multiple users might be the
``first come first served'' (FCFS) strategy in which
the system will serve the tenant who 
comes into the system first until it finds an optimal algorithm. The system then 
moves on to serve the next user. 
This strategy incurs a terrible cumulative regret of order $T$.

\vspace{0.5em}
\noindent
{\bf (Example)} Intuitively, it is easy to see why the FCFS strategy fails. Consider
two users, each of which has the best possible
model quality $100$. Each user has three models:
\vspace{-0.5em}
\begin{align*}
\texttt{U1 = \{M1: 90, M2: 95, M3: 100\}},\\
\texttt{U2 = \{M1: 70, M2: 95, M3: 100\}}.
\end{align*}
Assume \texttt{U1} comes 
into the system slightly earlier
and the system decides to try \texttt{M1} 
for \texttt{U1} in the first round. Then,
in round $t=1$, \texttt{U1} incurs
regret $r_{t=1}^1 = 100 - 90 = 10$
and \texttt{U2} incurs regret $r_{t=1}^2 = 100$
as it does not have a model to use. 
In the second round, the FCFS strategy
would continue to serve \texttt{U1}.
Assume the system tried \texttt{M2}.
Then, in round $t=2$, \texttt{U1} incurs
regret $r_{t=2}^1 = 100 - 95 = 5$
and \texttt{U2} continues to
incur regret $r_{t=2}^2 = 100$ because
it still has not been served. The accumulative
regret among all users has already become
215 at round $t=2$.
On the other hand, if the system decides to
serve \texttt{U2} in the second round,
\texttt{U1} would incur
regret $r_{t=2}^1 = 10$, which is the
same as the first round,
and \texttt{U2} will
incur regret $r_{t=2}^2 = 100 - 70 = 30$.
In this case, the accumulative
regret among all users would become
150 at round $t=2$.

\vspace{0.5em}
This example shows the importance of
choosing the next user to serve. The goal of
\eml is to design an algorithm
that automatically manages resource
allocations among multiple users.

\vspace{-0.5em}
\paragraph*{Discussion: Relation to 
Multi-task Model Selection}
We compare our multi-tenant
regret with the most similar multi-task
regret recently
studied by Swersky et al.~\cite{Swersky2013}.
Swersky et al. focus on 
k-fold cross validation. They 
take advantage of the correlation between
folds by skipping the evaluations for certain folds for the same model. Here, each fold corresponds to 
one of our tenants. The
multi-task regret optimized 
is $\sum_{t=1}^T \sum_{i=1}^n f(x, i, t)$,
where $f(x, i, t)$ is the regret incurred
at time $t$ for fold $i$ after choosing
model $x$. The key difference
is that, in their setting, choosing a
better $x$ lowers the regret for {\em all}
users, while in our setting users
who do not get served continue to incur 
the same regret as before. 
Directly using their algorithms would 
cause a problem in \eml. Consider
a case when two users are strongly correlated. Swersky et al. would try to 
skip evaluating one user completely
simply because it does not provide much new information. \eml, on the other hand, must 
balance between two users as it cannot 
simply reuse the model trained 
for user $i$ to serve a different user $j$.
As a result, the \eml algorithm is also very different.

\subsection{Round-Robin GP-UCB}

We start from a very simple strategy. Instead of
serving users in an FCFS manner, what if
we serve them in a ``round robin'' way?
More precisely, with $n$ users, 
at round $t$ the system
chooses to serve one user: $i = t \mod n$.
When that user gets served, she runs one
iteration of her own GP-UCB algorithm. 
Intuitively, this strategy enforces
the absolute ``fairness'' among all users.

\vspace{-0.5em}
\paragraph*{Theoretical Analysis}
Surprisingly, the simple \rr strategy already
has a much better regret bound compared
to FCFS. The following theorem shows
the result. 

\vspace{-0.5em}
\begin{theorem}
  \label{thm:rr}
  Given $\delta\in (0,1)$, set $\beta_t^i = 2c^\ast \log\left[\frac{\pi^2 n K^\ast t^2}{6\delta}\right]$ with $c^\ast = \max_{i\in [n], k\in [K^i]}\{c^i_k\}$ and $K^\ast = \max_{i\in [n]}\{K^i\}$. Then the
  cumulative regret of \rr is bounded by
  \[
     R_T \leq \sqrt{n T}\sum_{i=1}^n \sqrt{I_i([T(i)])}
  \]
  with probability at least $1-\delta$,
  where
  \begin{align*}
    I_i([T(i)]) & = \frac{8 (c^\ast)^2 \beta^\ast}{c_\ast \log(1 + (\sigma^\ast)^{-2})}\\
    &\qquad \sum_{t\in T(i)} \log\left(1 + (\sigma^i)^{-2}(\sigma^i_{t-1}(a^i_{t}))^2\right),
  \end{align*}
  with $c_\ast = \min_{i\in [n], k\in [K^i]}\{c^i_k\}$,  $\beta^\ast = 2c^\ast \log\left[\frac{\pi^2 n K^\ast T^2}{6\delta}\right]$, $\sigma^\ast= \max_{i\in [n]}\{\sigma^i\}$, and $T(i)$ is the set of time points when tenant $i$ is picked up to time $T$, i.e.,
  $T(i) = \{s: 1\le s\le T, I_s = i\}$.
\end{theorem}

In the \rr case, each tenant is picked for the same number of rounds, that is, all $|T(i)|$'s are roughly the same, bounded by $\lceil T/n\rceil$. Then $I_i([T(i)])$ can be uniformly upper bounded by $\beta^\ast \log (T/n)$ for commonly used covariance (kernel) matrices; see Theorem 5 in \cite{SrinivasKKS10}. Consequently, the cumulative regret is upper bounded, up to a constant, by
\begin{equation}
n^{1.5}\sqrt{\beta^\ast T \log |T(i)|} = n^{1.5}\sqrt{\beta^\ast T \log {T\over n}}.
\label{eq:RR_simple_bound}
\end{equation}
Thus, when $T \rightarrow \infty$, we have $R_T'/T < R_T/T \rightarrow 0$.

\vspace{-0.5em}
\paragraph*{Practical Considerations}
From a theoretical point of view,
\rr only has slightly worse theoretical bound
than the \greedy scheduler used by \eml.
Practically, it outperforms FCFS 
significantly. However, there is
still room to improve. 

The enforcement of absolute fairness
with a \rr strategy means that it may
waste resources on users who already have
reached their optimal solution. For example,
consider two users \texttt{U1}
and \texttt{U2} that both have the best
possible quality 100. Assume that \texttt{U1}
has already reached accuracy 99 while 
\texttt{U2} is still at 70. \rr
will still schedule both users
fairly. However, it is clear in this case
that a better scheduler would put more
resources on \texttt{U2}. This requires
us to automatically balance between
different users and to disproportionately 
serve users who the system believes
could contribute more to the ``global satisfaction''
of all users. However, reaching this
goal is not trivial. For each user,
we do not know the real ``best possible
quality'' beforehand. Thus, we need
to estimate the potential for 
improvement for each user from their
execution history and fuse these
estimations in a comparable and balanced
way across all users in the system.
This motivates
the novel \greedy algorithm we designed for \eml.

\begin{algorithm} [t!]                     
\small
\caption{\greedy, cost-oblivious, multi-tenant GP-UCB}          
\label{alg:mc-general}                           
\begin{algorithmic}[1]                    
   \REQUIRE GP prior $\{\mu_0^i\}_{i=1}^n$, $\{\sigma^i\}_{i=1}^n$, $\{\Sigma^i\}_{i=1}^n$, and $\delta\in (0,1)$
   \ENSURE Return the best algorithm among all algorithms in $a_{[1:T]}^i$ for all users $[n]$.
   \FOR{$i=1,...n$}
       \STATE Initialize $\sigma_0^i = \text{diag}(\Sigma^i)$ and $t_i=1$
       \STATE Run one step of GP-UCB (i.e., lines 3 to 5 of Algorithm~\ref{alg:gp-ucb}) for the user $i$ to obtain $\sigma^i_{t_i}$ and $\mu^i_{t_i}$
   \ENDFOR
   \FOR {$t=1,...,T$}
        \STATE Refine regrets for users $i\in[n]$ with new observations
        \[
   \tilde{\sigma}^i_{t_i-1} = \min\left\{B_{t_i-1}(a^i_{t_i-1}), \min_{t'<t_i-1} y^i_{t'}+ \tilde{\sigma}^{i}_{t'} \right\} - y^i_{t_i-1} 
   \]
   where $
   B_t(k)=\mu_{t-1}(k) + \sqrt{\beta_t}\sigma_{t-1}(k)$.
      \STATE Decide the candidate set
   \[
     V_t := \left\{ i\in [n]: 
         \tilde{\sigma}^i_{t_i-1} \ge \frac{1}{n}
       \sum_{i=1}^n 
         \tilde{\sigma}^i_{t_i-1}\right\}
   \]
   \STATE Select a user (using any rule) from $V_t$ indexed by $j$

   \STATE  Update $\beta_{t_j}^j$ by
   \[
     \beta_{t_j}^j = \log (K^j t_j^2/\delta)
   \]
   \STATE Select the best algorithm for the user $j$
   \[
     a^j_{t_j} = \argmax_{k\in [K^j]}\mu^j_{{t_j}-1}(k) + \sqrt{\beta^j_{t_j}}\sigma^j_{t_j-1}(k)
     \]
     \STATE Observe $y^j_{t_j}(a^j_{t_j})$

     \STATE Update $\sigma^j_{t_j}$ and $\mu^j_{t_j}$ (see lines 4 and 5 in Algorithm~\ref{alg:gp-ucb})
        \STATE Increase the count of the user $j$ by $t_j \leftarrow t_j + 1$
   \ENDFOR
\end{algorithmic}
\end{algorithm}

\subsection{Greedy GP-UCB}

We now describe the \greedy algorithm, an 
algorithm that improves over \rr by 
dynamically allocating resources.

\paragraph*{Intuition}
The intuition behind \greedy is to have
each user run their own GP-UCB. At round
$t$, when the system tries to decide which
user to serve, it estimates the
``potential'' of each user regarding how much each
user could further benefit from another
model-selection step. The system then
chooses a user with high-enough potential.
The technical difficulty is how to
estimate the potential for each user 
in a way that is comparable among all users.

\paragraph*{The Greedy Algorithm}

Algorithm~\ref{alg:mc-general} shows the details 
of the \greedy algorithm. For simplicity,
these are shown only for the cost-oblivious version of the algorithm. The cost-aware version simply replaces all occurrences of $\sqrt{\beta}$ by $\sqrt{\beta/c}$, where $c$ is the corresponding cost.
The \greedy algorithm consists of two phases.
In the first phase (lines 6 to 8), we determine which user to schedule next (i.e., the \emph{user-picking} phase).
In the second phase (lines 9 to 12), we determine which model to run for this user (i.e., the \emph{model-picking} phase).
The model-picking phase is straightforward. We choose the model with respect to the single-tenant GP-UCB criterion. (Compare lines 9 to 12 in Algorithm~\ref{alg:mc-general} to lines 3 to 5 in Algorithm~\ref{alg:gp-ucb}.)
The user-picking phase is more sophisticated.

One idea for user selection is to compare the best models from different users and pick ``the best of the best.''
However, in what sense are the best models comparable?
To understand the subtlety here, consider two models \texttt{M1} and \texttt{M2} from two users.
Suppose that, at a certain time step $t$, the mean quality of \texttt{M1} and \texttt{M2} is 90 and 70, respectively.
Moreover, assume that the Gaussian variances of \texttt{M1} and \texttt{M2} are the same.
The UCB criterion will then favor \texttt{M1} over \texttt{M2}.
In the single-tenant case, this makes perfect sense: \texttt{M1} is a clear win over \texttt{M2}.
In the multi-tenant case, this is perhaps indefinite because it is possible that \texttt{M1} is working on an easy data set whereas \texttt{M2} is working on a hard one.
Therefore, the mean quality in the UCB criterion (i.e., the $\mu$ part in the single-tenant GP-UCB algorithm) is not a reliable indicator of the potential model improvement when comparing different users.
Based on this observation, we choose to omit the mean quality in the UCB criterion and focus on the observed variance.

This leads to the specific user-selection strategy illustrated in lines 6 to 8 of Algorithm~\ref{alg:mc-general}.
In line 6, we first compute more accurate regrets for users with new observations.
Clearly, line 6 represents a recurrence relation on the \emph{empirical} confidence bounds $y^i_{t}+ \tilde{\sigma}^{i}_{t}$ that replace the means in the upper confidence bounds by the actual observations $y$.
Assuming that user $i$ was scheduled at time $t-1$, the empirical confidence bound for user $i$ after time $t-1$ (i.e., at time $t$) is either the updated upper confidence bound $B_{t-1}$
or the minimum empirical confidence bound before time $t$, whichever is smaller.
Intuitively, the empirical confidence bound tries to tighten the upper confidence bound by utilizing the observed reward more directly.
(The UCB criterion merely uses the observations to update the parameters of the Gaussian Process.)
In lines 7 and 8, we further use the empirical variances $\tilde{\sigma}^2$ of the users to determine which user to schedule next.
Specifically, we first identify a set of candidate users whose empirical confidence bounds are above the average and then pick one user from the candidates.

\paragraph*{Strategy for Line 8}
By choosing a user with a confidence bound above the average, we can reduce the time-averaging regret.
It is interesting that the regret bound remains the same regardless of the rule for picking a user from the candidates (line 8), though in practice a different rule may make a difference.
For example, picking the user with the maximum empirical variance may be better than randomly picking a user. In \eml, we 
use a rule that picks the user with the maximum
gap between the largest upper confidence bound
and the best accuracy so far.
Nevertheless, the existence of an optimal rule for deciding on the best candidate user in the practical sense remains as an open question.

\paragraph*{Theoretical Analysis} We
can prove the following regret bound
for the \greedy algorithm.

\vspace{-0.5em}
\begin{theorem}
  \label{thm:multi-cost}
  Given $\delta\in (0,1)$, set $\beta_t^i = 2c^\ast \log\left[ \frac{\pi^2 n K^\ast t^2}{6\delta} \right]$. Then the
  cumulative regret of \greedy is bounded  by
  \[
    R_T \leq n \sqrt{T}\sqrt{\sum_{i=1}^n I_i([T(i)])}
  \]
  with probability at least $1-\delta$,
  where
  \begin{align*}
    I_i([T(i)]) & =  \frac{4c^\ast \beta^\ast}{\log(1 + (\sigma^\ast)^{-2})}\\
    &\qquad \sum_{t\in T(i)} \log\left(1 + (\sigma^i)^{-2}(\sigma^i_{t-1}(a^i_{t}))^2\right),
  \end{align*}
  with $c^\ast$, $K^\ast$, $\sigma^\ast$, $\beta^\ast$, and $T(i)$ defined in Theorem~\ref{thm:rr}.
\end{theorem}

$I_i([T(i)])$ is proportional to the information gain for each user $i$. In particular, if the kernel function or covariance matrix is linear (see
  Theorem 5 in \cite{SrinivasKKS10}), then for each $i\in [n]$, the
  order of $I_i([T(i)])$ is at most $\beta^\ast \log(|T(i)|)$. Since
  \[
    \sum_{i=1}^n \log(|T(i)|) = \log \prod_{i=1}^n |T(i)|,
    \]
  with the constraint
    $\sum_{i=1}^n|T(i)| = T,$
  we have
  \[
    \sum_{i=1}^n \log(|T(i)|) \le n \log \left(\frac{T}{n}\right).
  \]
  In this case, the total regret is bounded (up to some constant) by
  \[
 n^{3/2}\sqrt{\beta^\ast T\sum_{i=1}^n \log(|T(i)|)}  \leq \underbrace{n^{3/2}\sqrt{\beta^\ast T \log\left( \frac{T}{n}\right)}}_{\text{the regret for RR, see \eqref{eq:RR_simple_bound}}}.
  \]
For two other popular kernels -- the squared exponential and the Mat\'{e}rn kernel -- we can also get a bound that is sublinear in $T$, and thus $R_T/T \to 0$ as $T\to\infty$ (Section 5.2 in \cite{SrinivasKKS10}). We see that the regret of \greedy is slightly better than the one of \rr. 

\vspace{-0.5em}
\subsection{A Hybrid Approach}

One problem of Algorithm~\ref{alg:mc-general} is that it may enter a \emph{freezing stage} and never step out.
That is, after a certain time step (usually at the very end of running the algorithm), the candidate set of users will remain stable.
The algorithm will stick to these users forever and therefore make no further progress if the optimal models for these users have been found.
The reason for this phenomenon is that the empirical variance, though improved over the Gaussian variance in the UCB criterion, is still an estimated bound rather than the true gap between the observed and optimal model quality.
When the observed model quality is close to the optimal quality, this estimated bound is no longer reliable as an indicator of the true gap.
Consequently, the empirical variances for the users remain almost constant, which results in a stable candidate set.

To overcome this problem, we further propose a hybrid approach.
When we notice that the candidate set remains unchanged and the overall regret does not drop for $s$ steps,
we know that the algorithm has entered the freezing stage.
We then switch to the round-robin user-selection strategy so that the algorithm can leave the freezing stage and make progress on the other users.
We used this hybrid approach in our experimental evaluation (Section~\ref{sec:experiments}), where we set $s=10$.
As we have just discussed, this hybrid approach observes the same regret bound as Algorithm~\ref{alg:mc-general} because using a round-robin user-selection strategy instead does not change the regret bound.

\vspace{-0.5em}
\subsection{Discussion on Limitations} \label{sec:limitations}

As the first attempt
at resolving the multi-tenant model-selection problem,
this paper has the following limitations.
From the theoretical perspective, the regret bound
is not yet theoretically optimal.
From the practical perspective,
the current framework only supports the case where
the whole GPU pool is treated as a single device.
With our current scale, this is not
a problem as our deep learning
subsystem still achieves significant speed up
in our setup
(all machines are connected with InfiniBand,
all communications are in low precision~\cite{zhang2016zipml},
and we tune learning rate following Goyal et al.~\cite{2017arXiv170602677G}).
However, as our service grows
rapidly, this nice scalability will soon disappear. 
We will need to extend our current framework
such that it is aware of multiple devices in the near future.
Another limitation is that our
analysis focuses on GP-UCB and it is not clear
how to integrate other algorithms such as
GP-EI~\cite{SnoekLA12} and GP-PI~\cite{Kushner1964} into a multi-tenant framework.
Last, we define the global
satisfaction of all users as the sum 
of their regrets. It is not clear how to
integrate hard rules such as the ``each user's deadline'' 
and design
algorithms for other aggregation functions.

\section{Experiments}\label{sec:experiments}

We present the experimental results of 
\eml for the real and synthetic data sets.
On the real data set, we validate that
\eml is able to provide
better global user experiences on one
real service we are providing to our users. 
We then use five synthetic data sets
to better understand the multi-tenant model-selection subsystem inside \eml. We validate
that each of the technical contributions 
involved in the design of our 
multi-tenant model-selection subsystem
results in a significant speedup
over strong baselines.

\subsection{Data Sets}

We now describe the six data sets we used
in the experiment. Figure~\ref{tab:datasets} summarizes these data sets.
Each data set used in our experiment 
contains a set of users and machine learning models. For each (user, model)
pair, there are two measurements
associated with it, namely (1) quality
(accuracy) and (2) cost
(execution time).

\vspace{0.5em}
\noindent
{\bf (Real Quality, Real Cost)} The 
\textsc{DeepLearning} data set was collected
from the \eml log of 22 users running image classification tasks. Each
user corresponds to a data set and
\eml uses eight models for
each data set:
NIN, GoogLeNet, ResNet-50, AlexNet, BN-AlexNet, ResNet-18, VGG-16, and SqueezeNet.
Each (user, model) pair is trained with
an Adam optimizer~\cite{Kingma2014}. The
system automatically grid-searches 
the initial learning rate in
\{0.1, 0.01, 0.001, 0.0001\} and
runs each setting for 100 epochs.

\vspace{0.5em}
Unfortunately, image classification 
is the only \eml service so far that
has enough users
to be used as a benchmark data set.
To further understand the robustness
of \eml and understand the
behavior of each of our technical
contributions, we used a set
of synthetic data sets described
as follows.

\vspace{0.5em}
\noindent
{\bf (Real Quality, Synthetic Cost)} 
\textsc{179Classifier}
is a data set with real quality 
but synthetic costs. It contains
121 users and 179 models obtained from
Delgado et al.~\cite{DelgadoCBA14}
on benchmarking different machine-
learning models on UCI data sets.
We use each data set as a user.
Because the original paper does not
report the training time, we 
generate synthetic costs from the uniform distribution $\mathcal{U}(0, 1)$.

\vspace{0.5em}
\noindent
{\bf (Synthetic Quality, Synthetic Cost)}
We further generate a family of 
synthetic data sets with a 
synthetic data generator that generates
synthetic quality and synthetic cost.

For $N$ users and $M$ models, the synthetic
data generator contains a generative model 
for the quality $x_{i,j}$ of a 
model $j\in[K]$ for a user $i\in[N]$.
We consider the following two factors that can affect $x_{i,j}$.

\noindent
{\bf 1. User baseline quality} $b_i$: Different 
users have different degrees of difficulty associated with their tasks --- we can achieve higher accuracy on some tasks than on others.
For a user $i$, $b_i$ then describes how difficult the corresponding task is.
In other words, $b_i$ characterizes the inherent difficulty of $i$ and the final model quality $x_{i,j}$ is modeled as a fluctuation around $b_i$ (for the model $j$). We simply draw the $b_i$s from a normal distribution $\mathcal{N}(\mu_b,\sigma_b^2)$.

\noindent
{\bf 2. Model quality variation} $m_j$: 
We use $m_j$ to denote the fluctuation of model $j$
over the baseline quality. The generative model for 
$m_j$s needs to capture this model correlation.
Specifically, for each model $j$, we first assign a 
``hidden feature'' by drawing
from $f(j)\sim\mathcal{U}(0,1)$.
Then, we define the covariance between 
two models $j$ and $j'$ as 
$\Sigma_M[j, j']=\exp \left\{-\frac{(f(j) - f(j'))^2}{\sigma_M^2} \right\}$. We sample
for each user $i$: $
[m_1,...,m_K] \sim \mathcal{N}(0, \Sigma_M)$.

We combine the above two factors and 
calculate the model quality as $x_{i,j} = b_i + \alpha\cdot m_j$. Each synthetic data
set is specified by two hyperparameters:
$\sigma_M$ and $\alpha$. $\sigma_M$ captures
the strength of the model correlation, and $\alpha$
captures the weight of the model correlation in
the final quality. We vary these two
parameters and generate the four data sets
($\SYN(\sigma_M, \alpha)$) shown in Figure~\ref{tab:datasets}.

\begin{figure}
\centering
\small
\begin{tabular}{c | r r | c c c}
\hline
{\bf Dataset} & {\bf \# Users} & {\bf \# Models} & {\bf Quality} & {\bf Cost} \\
\hline
\textsc{DeepLearning} & 22 & 8 & Real & Real \\
\textsc{179Classifier} & 121 & 179 & Real & Synthetic \\
\hline
\textsc{SYN(0.01,0.1)} & 200 & 100 & Synthetic & Synthetic \\
\textsc{SYN(0.01,1.0)} & 200 & 100 & Synthetic & Synthetic \\
\textsc{SYN(0.5,0.1)} & 200 & 100 & Synthetic & Synthetic \\
\textsc{SYN(0.5,1.0)} & 200 & 100 & Synthetic & Synthetic \\
\hline
\end{tabular}
\vspace{-1em}
\caption{Statistics of Datasets}
\label{tab:datasets}
\vspace{-1em}
\end{figure}

\begin{figure}[t!]
\centering
\includegraphics[width=0.4\textwidth]{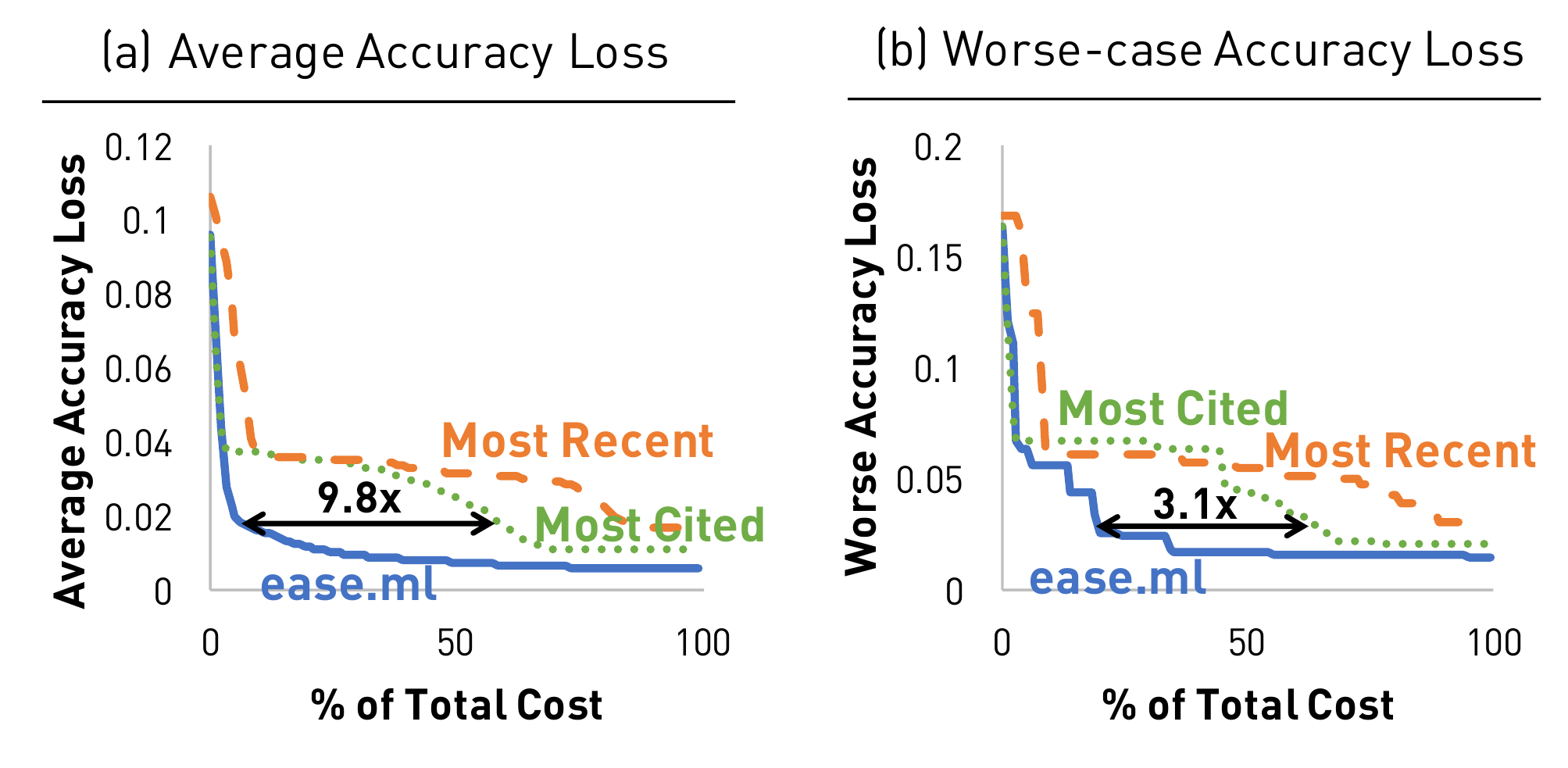}
\vspace{-1.5em}
\caption{End-to-end performance of
\texttt{ease.ml} on the
\textsc{DeepLearning} dataset
compared with two strategies
popularly used by our users 
without \texttt{ease.ml}. 
Each user uses one of
the two heuristics:
(1) most cited network first or
(2) most recently published network first. 
Different users are
scheduled with a round-robin scheduler.}
\label{fig:end-to-end}
\vspace{-1em}
\end{figure}

\subsection{End-to-End Performance of ease.ml}

We validate that \eml is able to achieve
a better global user experience than
an end-to-end system.

\vspace{-1em}
\paragraph*{Competitor Strategies}
We compare \eml with two strategies:
(1) \textsc{MostRecent} and (2) \textsc{MostCited}.
Both strategies use a round-robin scheduler
to choose the next user to serve. Inside
each user, it chooses the next model to train
by choosing the networks with the
most citations on Google Scholar or
published most recently. These two strategies
are the strategies that most of our users
were using before we provided them with \eml.

\vspace{-1em}
\paragraph*{Protocol} We run all
three strategies, namely (1) \eml,
(2) \textsc{MostRecent}, and (3) 
\textsc{MostCited}, on the 
data set \textsc{DeepLearning}.
We assume that all users enter the system 
at the same time and randomly sample
ten users as a testing set and the rest of the
users as a training set. For each
strategy, we run it for 10\% of the
total runtime of all models.
For \eml, all hyperparameters 
for GP-UCB are tuned by
maximizing the log-marginal-likelihood
as in scikit-learn.\footnote{\scriptsize
\url{http://scikit-learn.org/stable/modules/gaussian_process.html#gaussian-process}}
We repeat the experiment 50 times. 

\vspace{-1em}
\paragraph*{Metrics} We measure the performance
of all strategies in two ways.
For each of the 50 runs, we
measure the average of accuracy loss
among all users at a given time
point. Accuracy loss for each user
is defined as the gap between
the best possible accuracy among all
models and the best accuracy we
trained for the user so far. We
then measure the average across
all 50 runs as the first performance
measurement. Because
we treat ourselves as a ``service provider'',
we also care about the worst-case
performance and thus we measure the
worst-case accuracy loss 
across all 50 runs as another performance
measure.

\vspace{-1em}
\paragraph*{Results} Figure~\ref{fig:end-to-end}
illustrates the result.
We see that \eml outperforms the best of the two heuristics by up to 9.8$\times$ when comparing the average accuracy loss. The time spent on taking the average accuracy loss down from 0.1 to 0.02 of \textsc{MostCited} is about 9.8 times of (i.e., 8.8 times longer than) that of \eml.
In the worst case, \eml outperforms 
both competitors by up to 3.1$\times$.
We discuss in detail the reason for
such improvements together with other data
sets in the rest of this section.

\begin{figure}[t!]
\centering
\vspace{6em}
\includegraphics[width=0.5\textwidth]{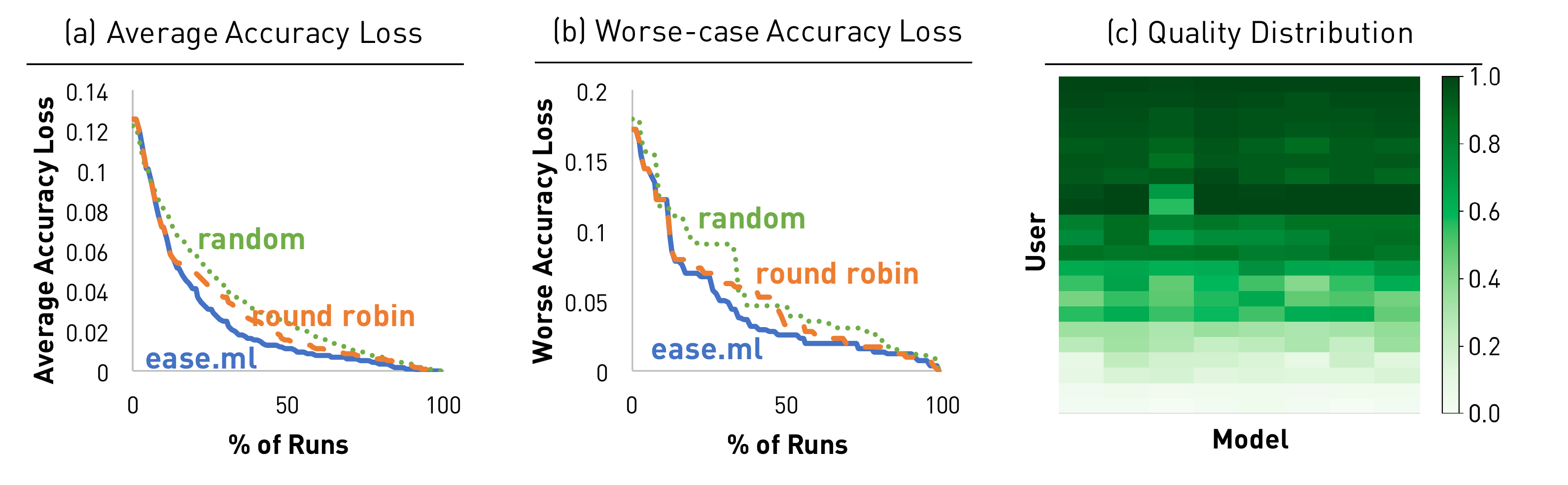}\\
\vspace{-0.5em}
\textsc{DeepLearning}\\
\vspace{1.5em}
\includegraphics[width=0.5\textwidth]{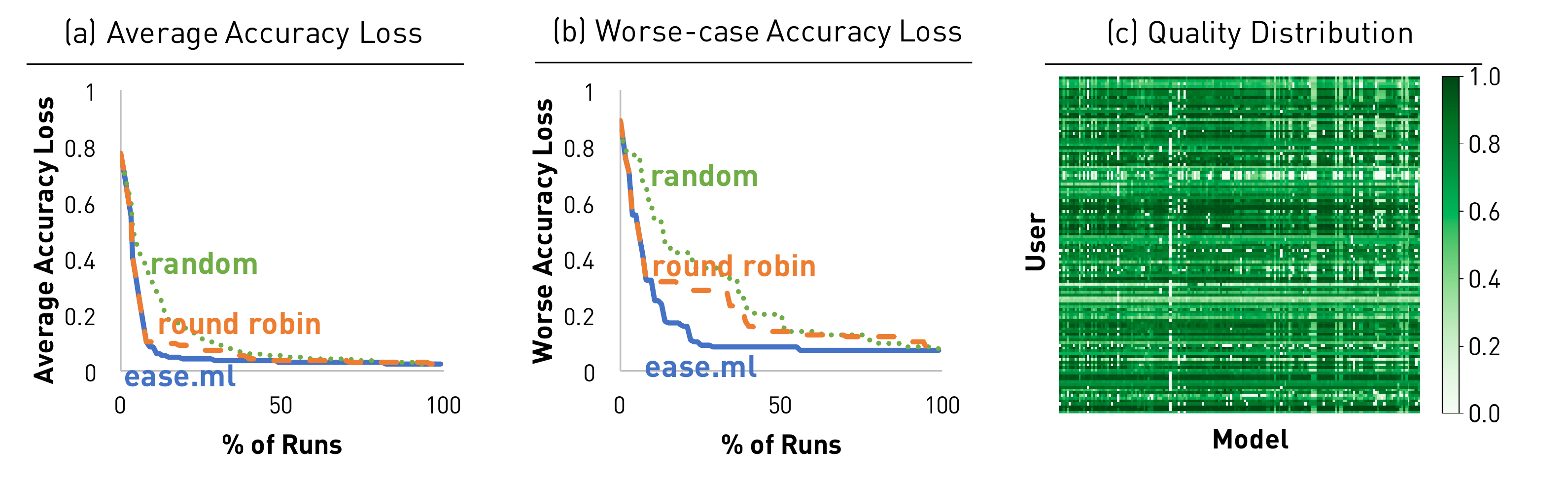}\\
\vspace{-0.5em}
\textsc{179Classifier}\\
\vspace{1.5em}
\includegraphics[width=0.5\textwidth]{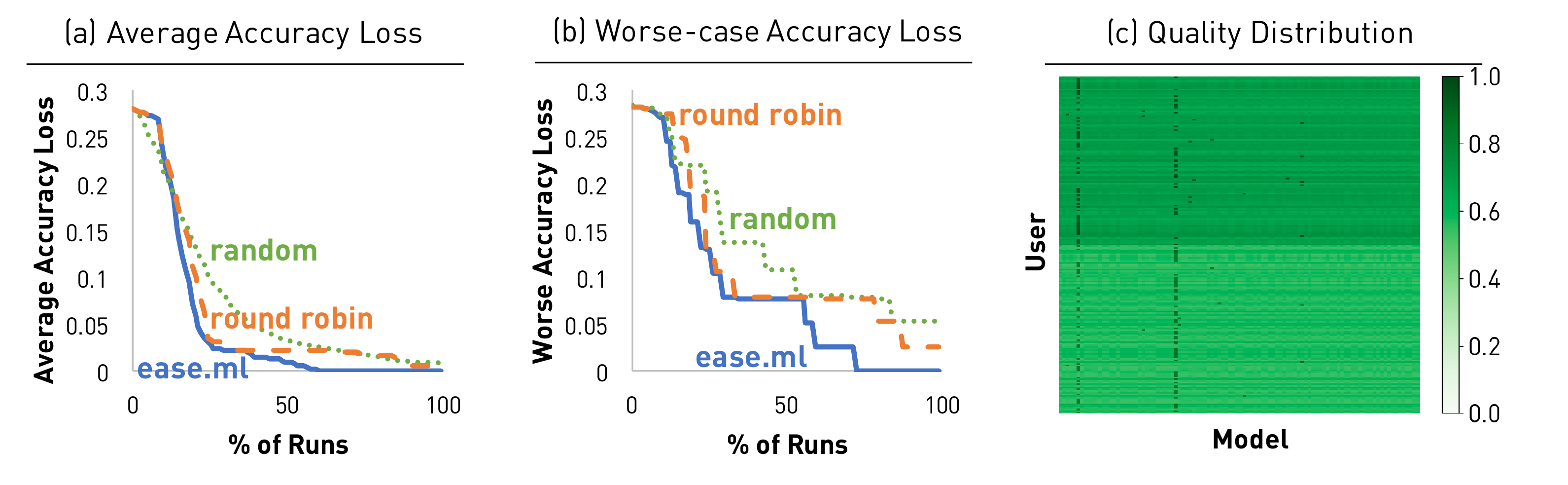}\\
\vspace{-0.5em}
\textsc{SYN(0.01, 0.1)}\\
\vspace{1.5em}
\includegraphics[width=0.5\textwidth]{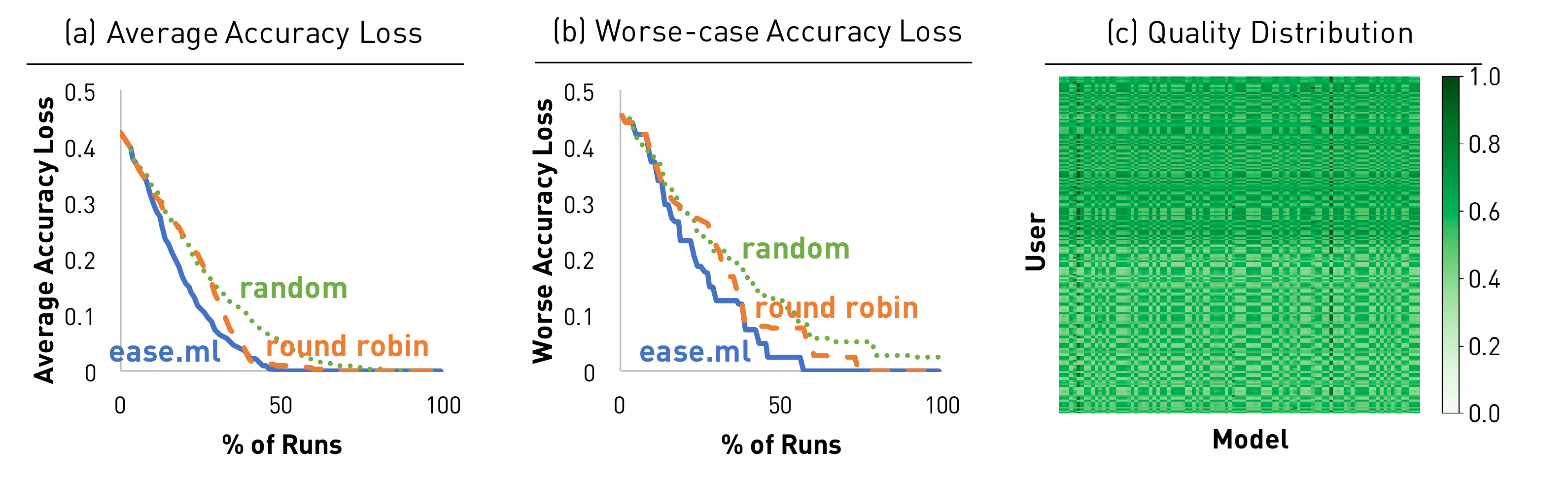}\\
\vspace{-0.5em}
\textsc{SYN(0.01, 1.0)}\\
\vspace{1.5em}
\includegraphics[width=0.5\textwidth]{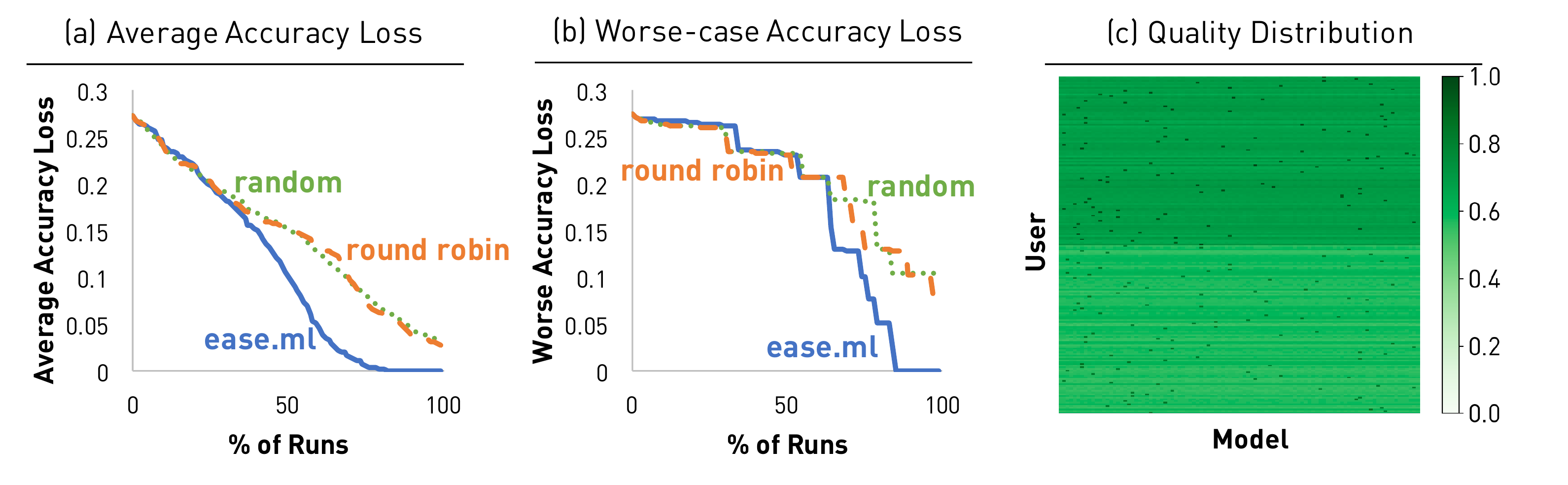}\\
\vspace{-0.5em}
\textsc{SYN(0.5, 0.1)}\\
\vspace{1.5em}
\includegraphics[width=0.5\textwidth]{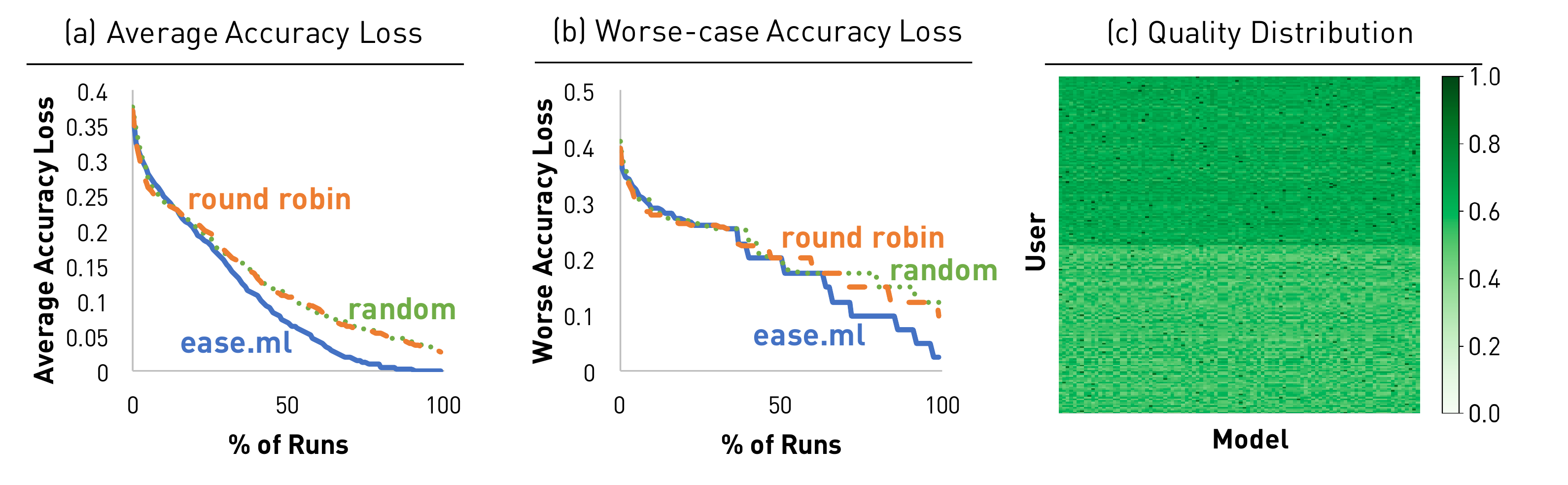}\\
\vspace{-0.5em}
\textsc{SYN(0.5, 1.0)}
\vspace{-0.5em}
\caption{Performance of the cost-oblivious case.}
\label{fig:exp:cost-oblivious}
\vspace{-2em}
\end{figure}

\begin{figure}[t!]
\centering
\vspace{6em}
\includegraphics[width=0.5\textwidth]{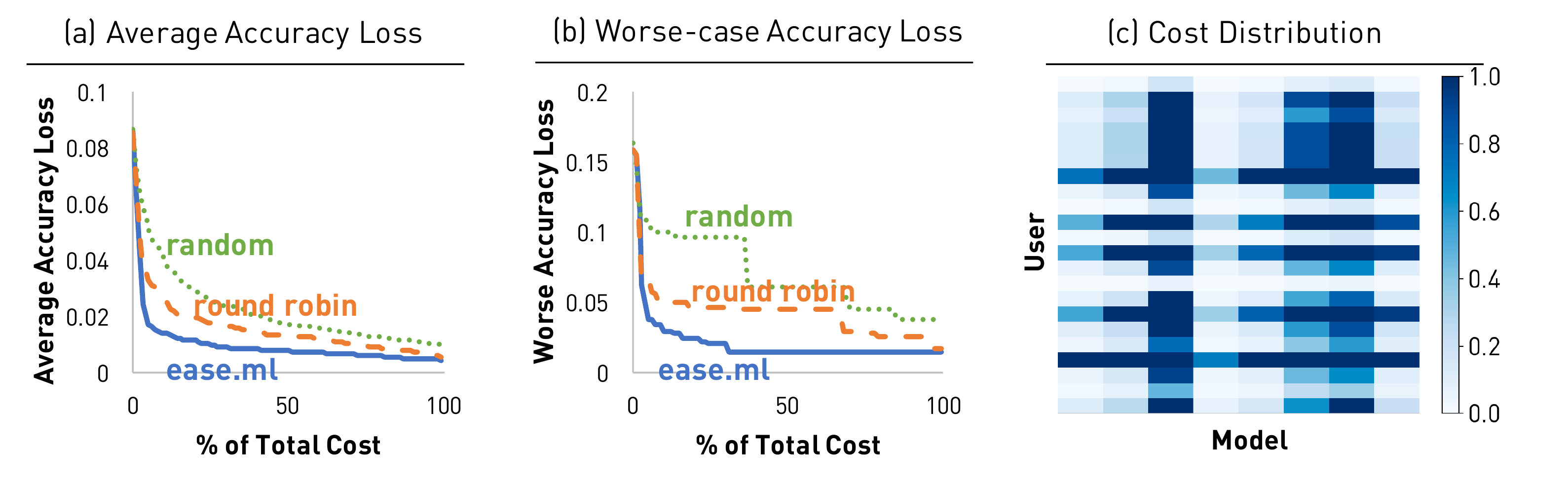}\\
\vspace{-0.5em}
\textsc{DeepLearning}\\
\vspace{1.5em}
\includegraphics[width=0.5\textwidth]{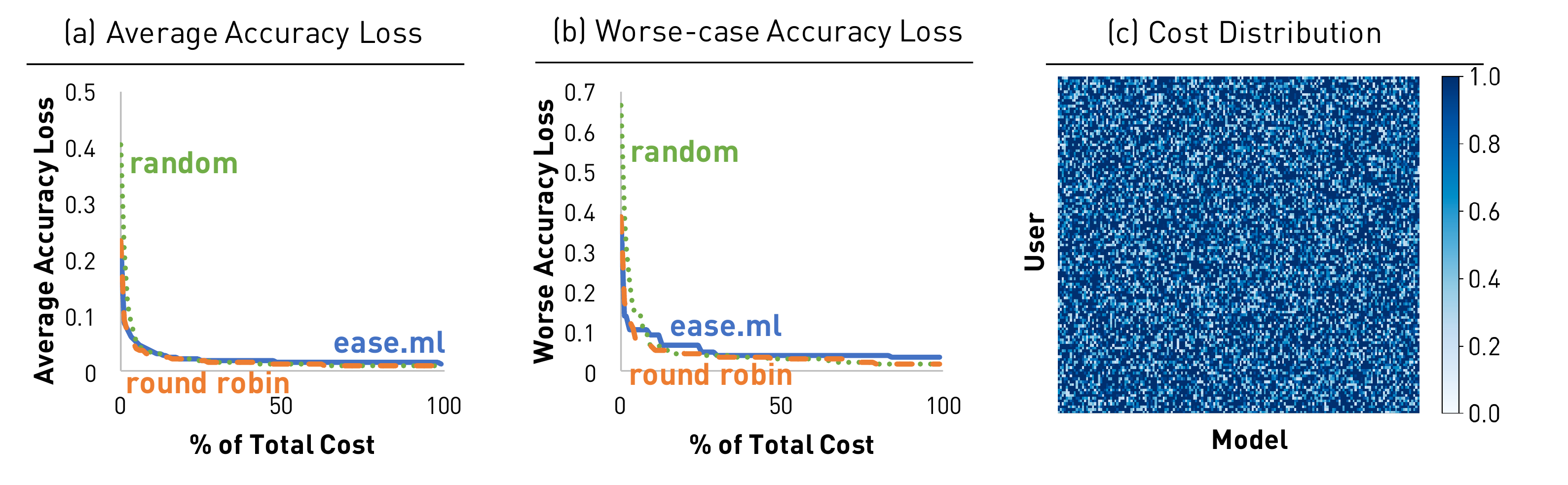}\\
\vspace{-0.5em}
\textsc{179Classifier}\\
\vspace{1.5em}
\includegraphics[width=0.5\textwidth]{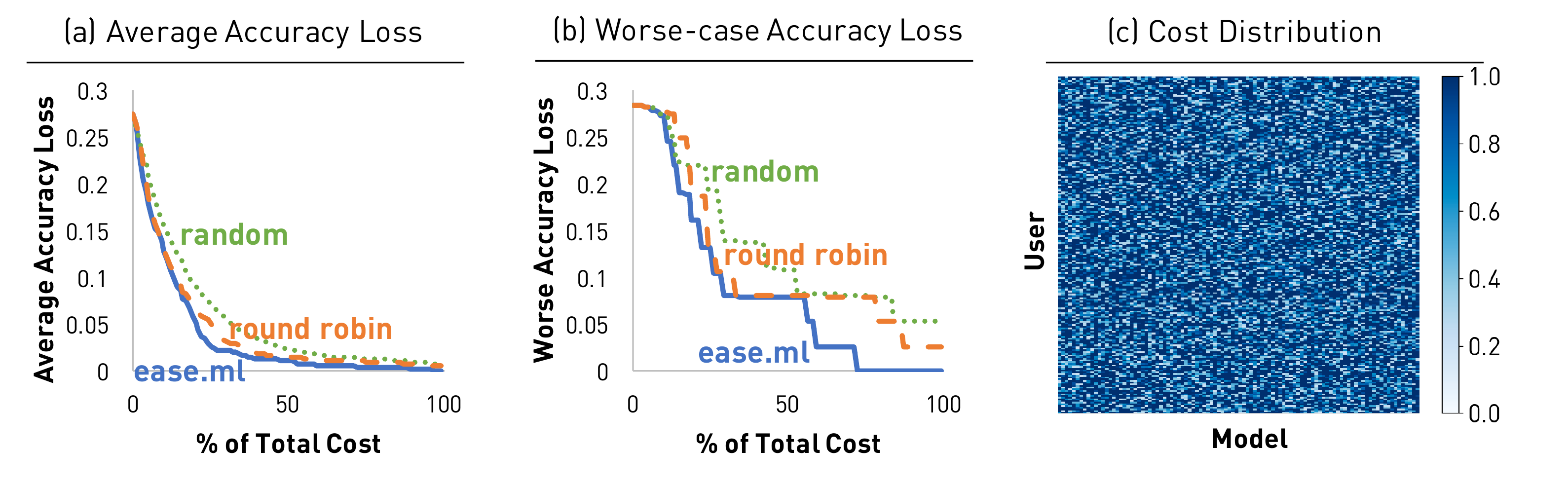}\\
\vspace{-0.5em}
\textsc{SYN(0.01, 0.1)}\\
\vspace{1.5em}
\includegraphics[width=0.5\textwidth]{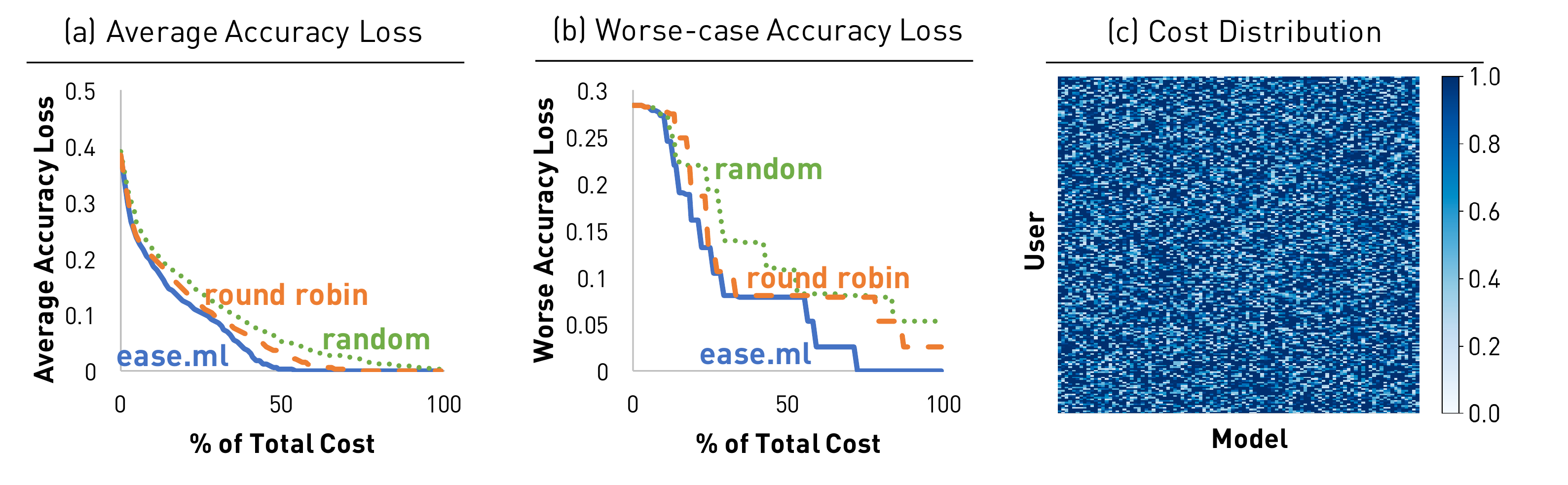}\\
\vspace{-0.5em}
\textsc{SYN(0.01, 1.0)}\\
\vspace{1.5em}
\includegraphics[width=0.5\textwidth]{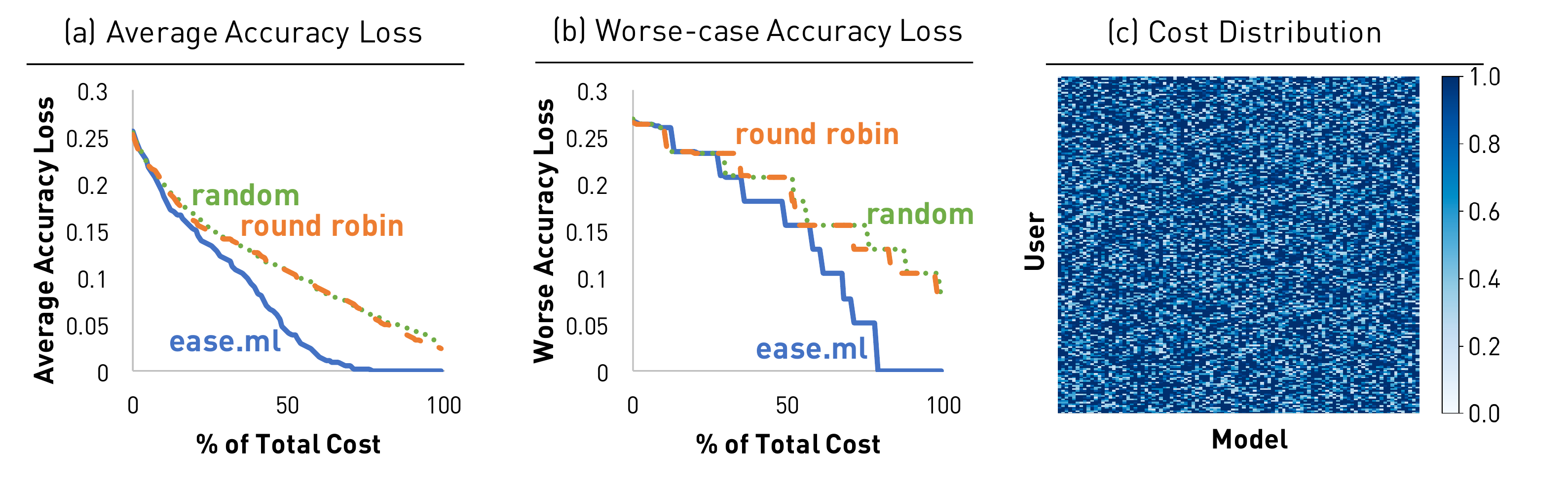}\\
\vspace{-0.5em}
\textsc{SYN(0.5, 0.1)}\\
\vspace{1.5em}
\includegraphics[width=0.5\textwidth]{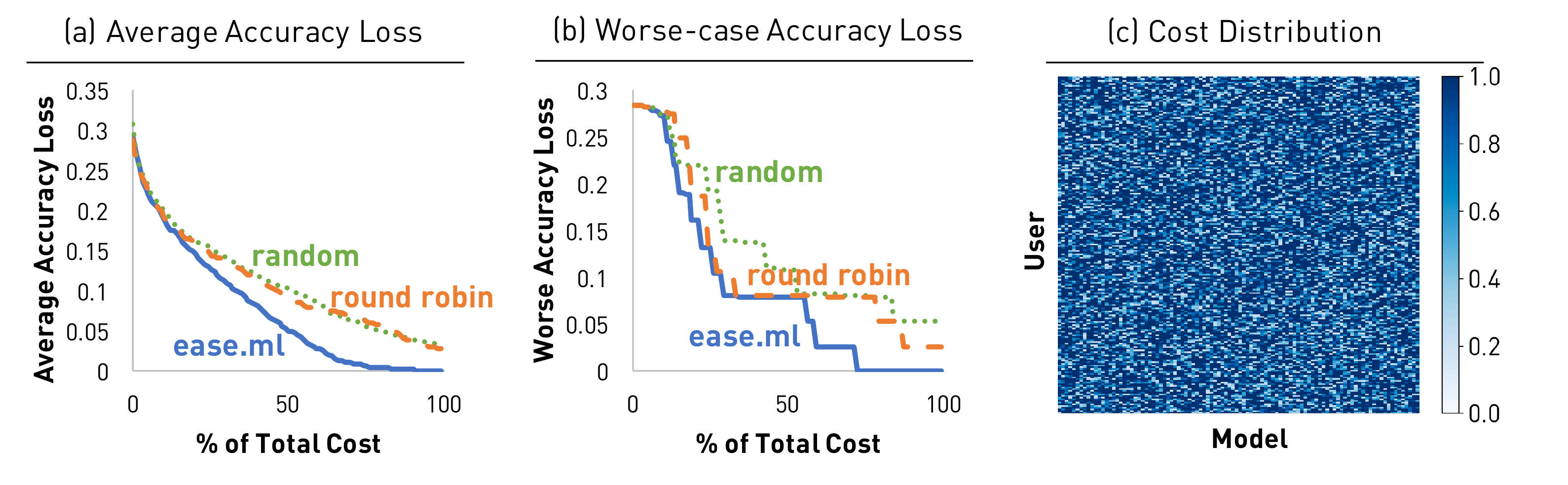}\\
\vspace{-0.5em}
\textsc{SYN(0.5, 1.0)}
\vspace{-0.5em}
\caption{Performance of the cost-aware case.}
\label{fig:exp:cost-aware}
\vspace{-2em}
\end{figure}

\begin{figure}[t!]
\centering
\includegraphics[width=0.5\textwidth]{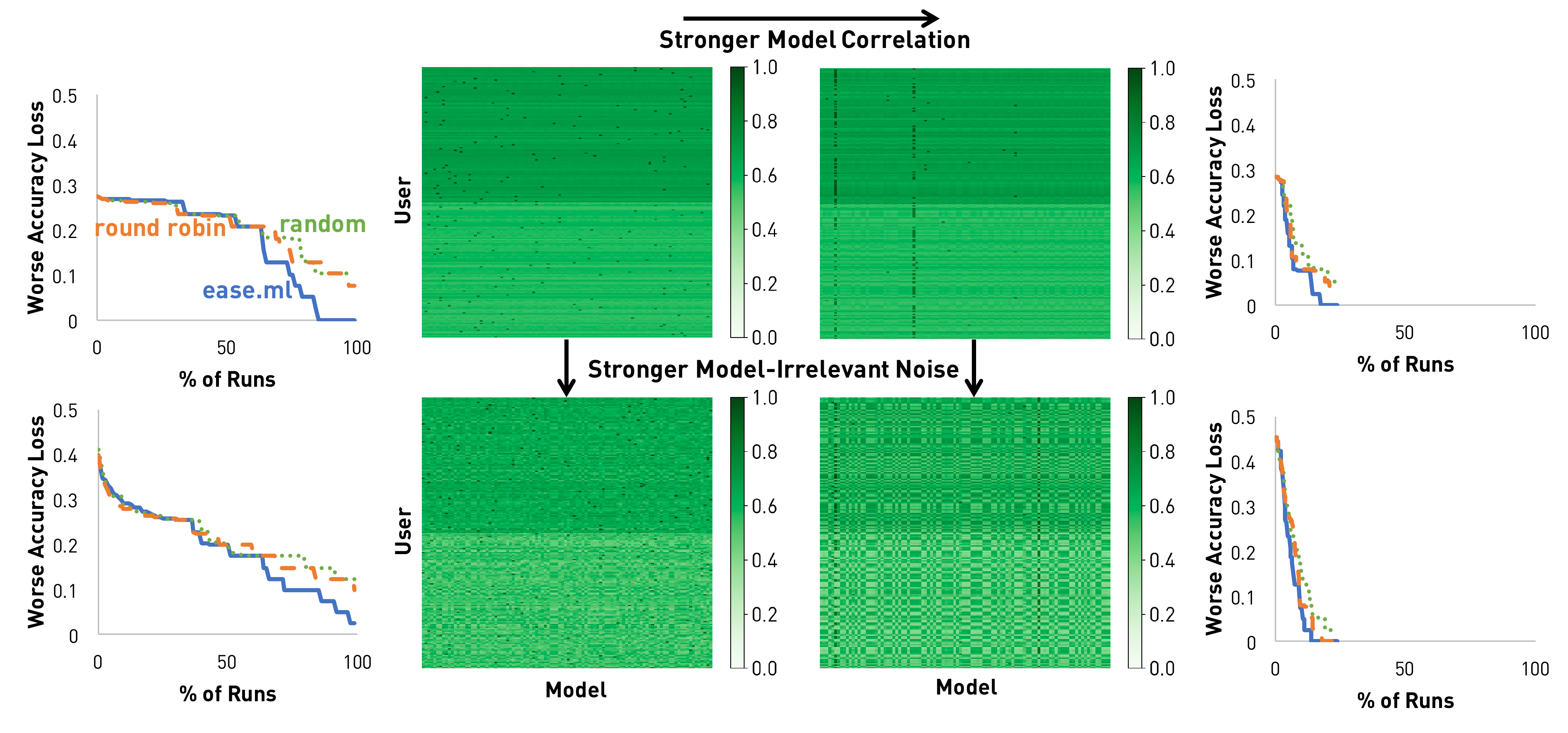}
\vspace{-3em}
\caption{The impact of model correlation and noise}
\label{fig:impact-model-correlation}
\vspace{-2em}
\end{figure}

\subsection{Multi-tenant Model Selection}

We validate that the end-to-end performance
improvement of \eml is brought about by the
multi-tenant, cost-aware model-selection
algorithm we propose in this paper.
We follow a similar protocol as the
end-to-end experiment with the following changes.

\vspace{-1em}
\paragraph*{Stronger Competitors} We compare
\eml with two even stronger baselines
than \textsc{MostCited} and \textsc{MostRecent} ---
instead of using heuristics to choose the
next models to run for each user, we
use GP-UCB for each user and
choose the next users to serve 
in a \textsc{RoundRobin} 
and \textsc{Random} way. In practice,
we observe \textsc{RoundRobin} outperforms both 
\textsc{MostCited} and \textsc{MostRecent}
on \textsc{DeepLearning}.

\vspace{-1em}
\paragraph*{Robustness to \# Users}

In addition to the end-to-end
protocol, we also ran
experiments with 50 users
in the testing set for data sets
with more than 100 users. The experiment
result is similar to the ten-user case.
Thus, we will only show the
results for the ten users.

\vspace{-0.5em}
\subsubsection{The Cost-oblivious, Multi-tenant Case}

We first evaluate the multi-tenant setting
when all systems are not cost-aware. In this
case, the performance is measured as in
\# runs instead of cost (execution time).
We run each system to allow it to
train 50\% of all available models.

\vspace{-1em}
\paragraph*{Results}
Figure~\ref{fig:exp:cost-oblivious} shows the 
results. Each row in Figure~\ref{fig:exp:cost-oblivious} corresponds
to one data set.
The first column shows the average accuracy loss (of the 50 runs of the same experiment) as the number of runs increases, whereas the second column shows the worst-case accuracy loss (among the 50 runs of the same experiment) as the number of runs increases.
The third column further shows the model-quality distribution of the underlying data set.

On the two datasets with real
quality, i.e., \textsc{DeepLearning} and 
\textsc{179Classifier}, we observe that the 
average and worst-case accuracy loss of \eml 
drops faster, up to 1.9$\times$,
compared to \textsc{Random} and 
\textsc{RoundRobin}. We also observe
similar results for all four synthetic data sets.
This shows that \eml reduces the total regret 
of all users faster than the other two baselines.

Compared with the end-to-end performance on 
\textsc{DeepLearning},
we observe that, in the cost-oblivious case, the performance
improvement of \eml is much smaller than
9.8$\times$. As we will see later, the 
large end-to-end improvement can only be
achieved when different models have different
execution costs, in which case 
prioritizing between users becomes more important.

In all experiments, \textsc{RoundRobin} 
slightly outperforms \textsc{Random}. This is not 
surprising as the only difference
between \textsc{Random} and \textsc{RoundRobin} can be explained as two
uniform samplings, one with and one without replacement.
\textsc{RoundRobin} 
outperforms \textsc{Random}
because it enforces {\em deterministic} 
fairness on all users, thought the performance gain is not much.

\vspace{-1em}
\paragraph*{Impact of Model Correlation}

One important factor that has an impact on
the performance of \eml is the strength
of correlation among the quality of all models.
Intuitively, the stronger the correlation,
the better \eml would perform because its
estimation on the quality of other models
becomes more accurate. We study the
impact of model correlation with the
four synthetic data sets
as shown in Figure~\ref{fig:impact-model-correlation}.
As we increase $\sigma_M$ from 0.01 to 0.5, the 
model correlation increases. As we reduce $\alpha$ from 1.0 to 0.1, the weight of the model correlation decreases, which implies that the impact of the model-irrelevant noise increases.

As illustrated in Figure~\ref{fig:impact-model-correlation}, the performance of the algorithms improves upon the stronger model correlation.
This is understandable. When model correlation becomes stronger, it is easier to distinguish good models from bad ones.
On the other hand, consider an extreme case when all the models are independent.
In such circumstance, an evaluation of one model cannot gain information about the others, which therefore requires more explorations before the overall picture of model performance becomes clear.
Meanwhile, dampening the impact of the model correlation has similar effects.

\vspace{-0.5em}
\subsubsection{The Cost-aware, Multi-tenant Case}

We now evaluate the multi-tenant setting when
all systems are aware of the cost,
the more realistic scenario that \eml is designed for.
For the \textsc{DeepLearning} data set, we use the real cost (execution time) of each model.
For the \textsc{179Classifier} data set and the four synthetic data sets, we generate costs randomly.

\vspace{-1em}
\paragraph*{Results}
Figure~\ref{fig:exp:cost-aware} shows the result.
The first two columns show the 
average accuracy loss and worst-case 
accuracy loss, respectively. 
The third column shows the cost distribution
of the underlying data set. All data sets
share the same quality distribution as 
the cost-oblivious case, as can be read
from Figure~\ref{fig:exp:cost-oblivious}.

The relative performance of the three 
algorithms is similar to the cost-oblivious case.
However, the improvement of
\eml becomes more significant. This is because
different costs magnify the differences
between different users, and thus
the multi-tenant algorithm in \eml
becomes more useful.

\begin{figure}[t!]
\centering
\includegraphics[width=0.3\textwidth]{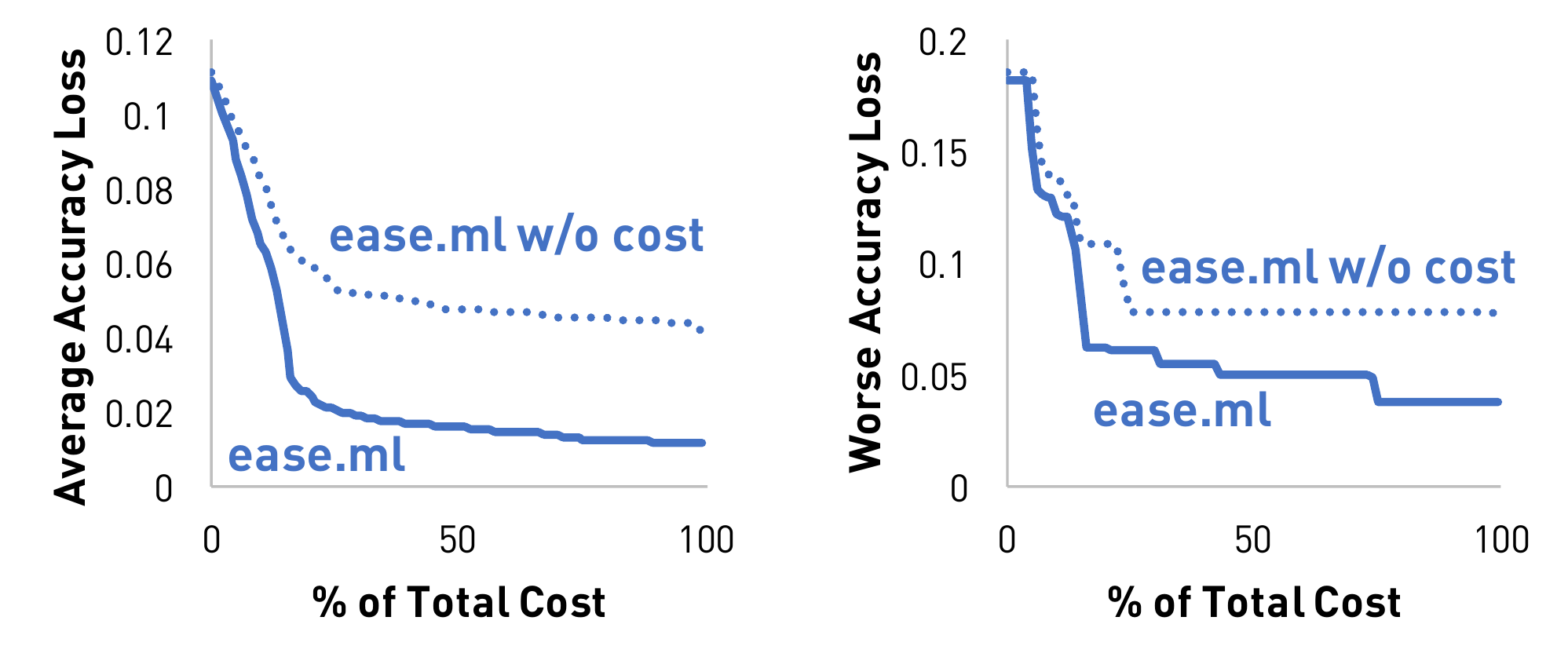}\\
\vspace{-0.75em}
\textsc{DeepLearning}\\
\vspace{-1em}
\caption{Lesion study: The impact of cost-awareness.}
\label{fig:impact-cost-aware}
\vskip -3ex
\end{figure}

\vspace{-1em}
\paragraph*{Impact of Cost-Awareness}
We validate the impact of using 
a cost-aware model-selection algorithm
for each user. We conduct a lesion
study by disabling the cost-aware
component in \eml (set $c_{i,j}=1$
for GP-UCB). Figure~\ref{fig:impact-cost-aware}
shows the result on the \textsc{DeepLearning} data set.
We see that considering the
execution cost of the model significantly
improves the performance of the system.
Combining the data distribution in
Figure~\ref{fig:exp:cost-oblivious}
and the cost distribution in 
Figure~\ref{fig:exp:cost-aware},
we see that this improvement
is reasonable --- models exist
that are significantly faster
on certain tasks and have a quality that
is only a little bit worse than the
best slower model. By integrating
cost into our algorithm, \eml is
able to automatically balance between
execution time and the increase in quality of the global
user experience.

\vspace{-1em}
\paragraph*{Impact of the Size of Training Data}

\begin{figure}[t!]
\centering
\includegraphics[width=0.3\textwidth]{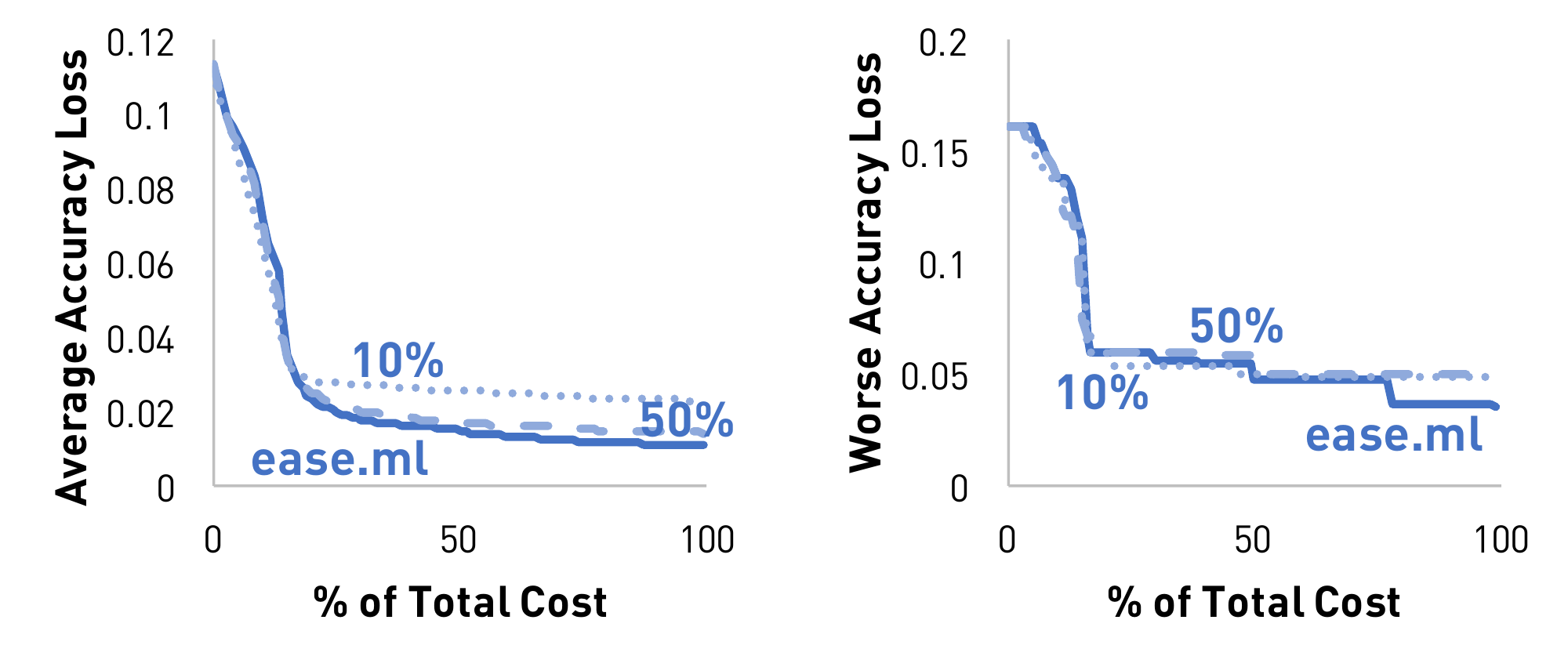}
\vspace{-2em}
\caption{The impact of training set size (cost aware).}
\label{fig:training:aware}
\vskip -1ex
\end{figure}

We validate that by providing \eml as
a service available to multiple users
and collecting the logs from all these users.
\eml is able to use this information to
help other users in the system. The
design of the algorithm in \eml achieves
this by calculating the kernel of the Guassian Process from the training set --- in other words,
the performance of a model on other users'
data sets defines the similarity (correlation)
between models. To validate the impact
of this kernel, we decrease the amount
of training data made available to
the kernel (10\%, 50\%, 100\%) and compare their performance. 
From Figure~\ref{fig:training:aware}
we see that having more models to 
calculate a kernel significantly improves
the performance of \eml. On the other hand,
we also observe the phenomenon of ``diminishing
return'' --- that is, using 50\% of the
training data results in similar performance
as using 100\% of the training data.

\begin{figure}[t!]
\centering
\includegraphics[width=0.15\textwidth]{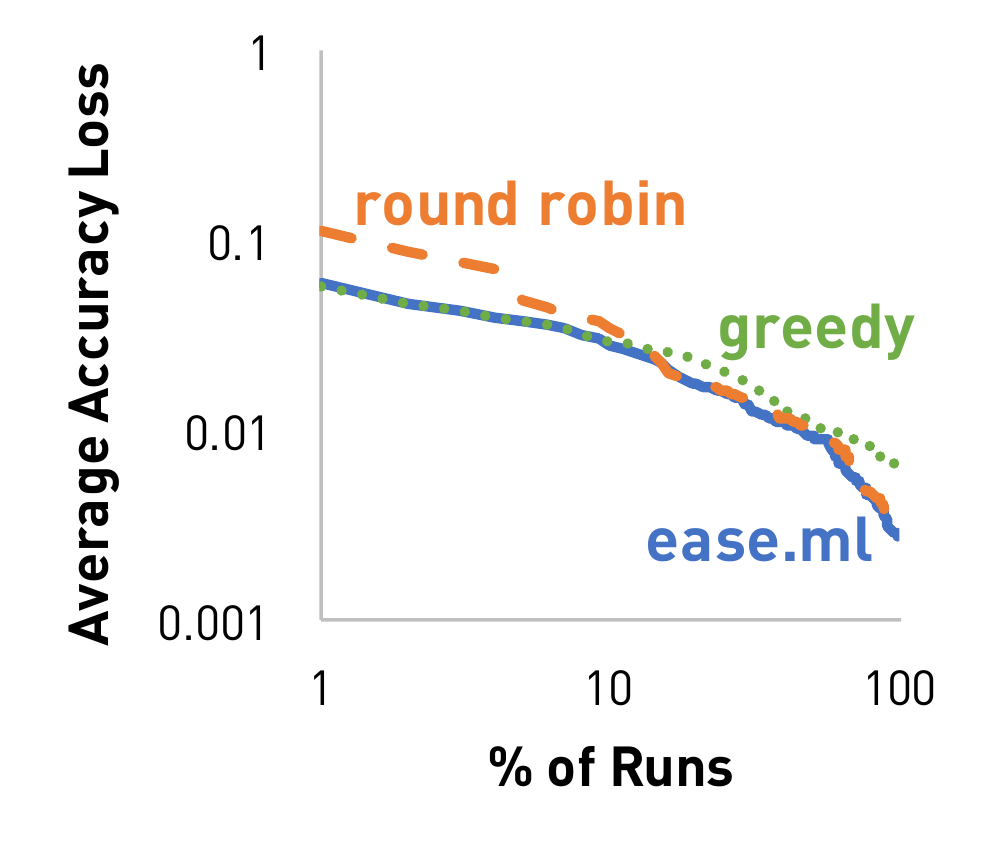}
\vspace{-2em}
\caption{Lesion study: The impact of hybrid execution.}
\label{fig:hybrid-exec}
\vspace{-2em}
\end{figure}

\vspace{-1em}
\paragraph*{Impact of Hybrid Execution}
We now validate the impact of the hybrid 
approach in \eml.
We disable the hybrid component and use each
of the two strategies for comparison: (1)
\textsc{Greedy} and (2) \textsc{RoundRobin}.
Figure~\ref{fig:hybrid-exec} shows the result
on the \textsc{179Classifier} data set
for the cost-oblivious case.

We observe that, while \textsc{Greedy} outperforms \textsc{RoundRobin} at the beginning, there is a crossover point as the algorithms proceed where \textsc{RoundRobin} becomes superior to \textsc{Greedy}.
Switching to \textsc{RoundRobin} after the crossover point makes the hybrid execution strategy the best among the three algorithms.
The reason that the crossover point exists
is because of the quality of the GP estimator:
As the execution proceeds, it is more
and more important for GP to have a good
estimation of the quality to choose
the next user to serve. When all users'
current accuracy is high enough, the modeling
error of treating the model-selection
problem as a Gaussian Process starts to become
ineligible. Thus, \textsc{RoundRobin}
works better for the second half
by serving users fairly.

\vspace{-1em}
\paragraph*{Discussion: Single- vs. Multi- Devices}
One decision we made in \eml that is reflected in  
our experiment protocol is that we are
treating all GPUs we have as {\em a single device}.
In our current setting (8 GPUs per machine while
all machines are connected by InfiniBand;
All data movements are in
low precision~\cite{zhang2016zipml}
and we follow Goyal et al), this does
not cause a problem because 
we can still get significant speed
up with all GPUs training the same model.
Consider a multi-device alternative
that allocates one GPU to each user.
Both alternatives require the similar 
amount of GPU-time to finish training. However,
the single-device strategy returns a model
faster for {\em some} users. In fact,
for our \textsc{DeepLearning} service,
the single-device option achieves lower 
accumulated regret among users 
than the multi-device alternative.
However, we realize that this design
is not fundamental and that it will be 
important in future work to make \eml aware of
multi-devices.


\vspace{-0.5em}
\section{Related Work}\label{sec:relatedwork}

From the system perspective, \eml is closely related to the
AutoML systems recently built by the database and machine learning communities.
Examples include Spark TuPAQ~\cite{Sparks2015}, Auto-WEKA~\cite{Kotthoff2017,Thornton2013}, 
Google Vizier~\cite{Golovin2017}, Spearmint~\cite{Snoek2012}, GPyOpt~\cite{gpyopt2016},
and Auto scikit-learn~\cite{Feurer2015}. Most of these systems
are built on state-of-the-art automatic model selection and 
hyperparameter tuning algorithms such as GP-UCB~\cite{SrinivasKKS10},
GP-PI~\cite{Kushner1964}, GP-EI~\cite{SnoekLA12}.
See Luo~\cite{Luo16} for an overview.
Compared to these systems, \eml is an AutoML system based on GP-UCB. However, the focus is
on the multi-tenant case in which multiple users share the same
infrastructure running machine learning workloads. To our best knowledge,
this is different from all existing model-selection systems.

\vspace{-1em}
\paragraph*{Multi-task Model Selection and Bayesian Optimization} 
The most relevant line of research is {\em multi-task model selection}
optimized for multiple concurrent single-tenant model selection tasks.
\eml contains a new multi-task model-selection algorithm. 

Compared to previous work, \eml focuses
on a different multi-tenant setting. For example, Swersky et al.~\cite{Swersky2013}
proposed different algorithms for two scenarios. 
Other than the cross validation application we discussed before,
it also consider cases when a cheaper classifier and an expensive classifier coexist.
Swersky et al.
try to query the cheaper classifier to get information
for the expensive one --- this is different from our objective
as their algorithm requires a ``primary user'' while in \eml
all users need to be treated equally. Similar observations
hold for previous work by Bardenet et al.~\cite{Bardenet2013} and Hutter et al.~\cite{Hutter2011}.
We designed \eml to optimize for an objective
that we generalized from our experience in serving our collaborators
and also designed a novel algorithm. 

\vspace{-1em}
\paragraph*{Multi-task Gaussian Process}
There has been work done in extending the Gaussian Process to the 
multi-task case. One technical example is the {\em intrinsic model
of coregionalization}~\cite{Goovaerts1997} that decomposes
a kernel with a Kronecker product. Most 
multi-task model-selection algorithms use some version of
a multi-task Gaussian Process as the underlying estimator.
\eml uses a simple multi-task Gaussian Process estimator in which we
do not consider the dependencies between users. One 
future direction will be to further integrate user correlations 
into \eml. 
Another related direction is parallel Gaussian Process in which multiple
processes are being evaluated instead of just one~\cite{Desautels2014}.
The key here is to balance the diversity of multiple samples, 
and integrating similar techniques
to extend \eml's resource model from a single device to multiple devices will be the subject of future work.

\vspace{-1em}
\paragraph*{Multi-tenant Clouds and Resource Management}
The focus of multi-tenancy on shared infrastructure is not new ---
in fact, it has been one of the classic topics intensively
studied by the database community. Examples include
sharing buffer pools~\cite{Narasayya2015} and CPUs~\cite{Das2013}
for multi-tenant relational ``databases-as-a-service''
and sharing semi-structured data sources~\cite{Bellare2013}.
\eml is inspired by such work but focuses on
multi-tenancy for machine learning.

\vspace{-1em}
\paragraph*{Declarative Machine Learning Clouds} 
There has been intensive study of declarative
machine learning systems~\cite{Abadi2016,Alexandrov2015,Boehm2016a,Hellerstein:2012:VLDB,Low2012,Meng2001,Sparks2017}. In \eml we 
study how to integrate automatic model selection into
a high-level abstraction and hope our result can be integrated
into other systems in the future.
Another related trend is the so-called ``machine learning
cloud.'' Examples include the Azure ML Studio
and Amazon ML. These services
provide a high-level interface and often support automatic model selection
and hyperparameter tuning. \eml is designed with the goal of
lowering the operating cost for these services 
by enabling multiple users sharing the same underlying 
infrastructure.

\vspace{-1em}
\section{Conclusion}\label{sec:conclusion}
\vspace{-0.5em}
In this paper, we presented \eml, a declarative machine learning system that automatically manages an infrastructure shared by multiple users.
We focused on studying one of the key technical problems in \eml, i.e., multi-tenant model-selection.
We gave the first formulation and proposed a solution that extends the well-known GP-UCB algorithm into the multi-tenant, cost-aware setting with theoretical guarantees.
Our experimental evaluation substantiates the effectiveness of \eml, which significantly outperforms popular heuristic approaches currently used in practice. 


\newpage

\balance
\bibliographystyle{abbrv}
\bibliography{vldb_sample}

\begin{thebibliography}{10}

\bibitem{Abadi2016}
M.~Abadi, A.~Agarwal, P.~Barham, E.~Brevdo, Z.~Chen, C.~Citro, G.~S. Corrado,
  A.~Davis, J.~Dean, M.~Devin, S.~Ghemawat, I.~Goodfellow, A.~Harp, G.~Irving,
  M.~Isard, Y.~Jia, R.~Jozefowicz, L.~Kaiser, M.~Kudlur, J.~Levenberg, D.~Mane,
  R.~Monga, S.~Moore, D.~Murray, C.~Olah, M.~Schuster, J.~Shlens, B.~Steiner,
  I.~Sutskever, K.~Talwar, P.~Tucker, V.~Vanhoucke, V.~Vasudevan, F.~Viegas,
  O.~Vinyals, P.~Warden, M.~Wattenberg, M.~Wicke, Y.~Yu, and X.~Zheng.
\newblock {TensorFlow: Large-Scale Machine Learning on Heterogeneous
  Distributed Systems}.
\newblock {\em ArXiv}, mar 2016.

\bibitem{Alexandrov2015}
A.~Alexandrov, A.~Kunft, A.~Katsifodimos, F.~Sch{\"{u}}ler, L.~Thamsen, O.~Kao,
  T.~Herb, and V.~Markl.
\newblock {Implicit Parallelism through Deep Language Embedding}.
\newblock In {\em Proceedings of the 2015 ACM SIGMOD International Conference
  on Management of Data - SIGMOD '15}, pages 47--61, New York, New York, USA,
  2015. ACM Press.

\bibitem{Bailis2017}
P.~Bailis, K.~Olukotun, C.~Re, and M.~Zaharia.
\newblock {Infrastructure for Usable Machine Learning: The Stanford DAWN
  Project}.
\newblock {\em arXiv}, may 2017.

\bibitem{Bardenet2013}
R.~Bardenet, M.~Brendel, B.~K{\'{e}}gl, and M.~Sebag.
\newblock {Collaborative hyperparameter tuning}.
\newblock In {\em ICML}, pages II--199. JMLR.org, 2013.

\bibitem{Bellare2013}
K.~Bellare, C.~Curino, A.~Machanavajihala, P.~Mika, M.~Rahurkar, and A.~Sane.
\newblock {WOO: a scalable and multi-tenant platform for continuous knowledge
  base synthesis}.
\newblock {\em Proceedings of the VLDB Endowment}, 6(11):1114--1125, aug 2013.

\bibitem{Binnig2016}
C.~Binnig, A.~Fekete, A.~Nandi, {Association for Computing Machinery},
  C.~{ACM-Sigmod International Conference on Management of Data (2016 : San
  Francisco}, and C.~{ACM SIGACT-SIGMOD-SIGART Symposium on Principles of
  Database Systems (2016 : San Francisco}.
\newblock {\em {Proceedings of the Workshop on Human-In-the-Loop Data
  Analytics}}.
\newblock ACM, 2016.

\bibitem{Boehm2016a}
M.~Boehm, A.~C. Surve, S.~Tatikonda, M.~W. Dusenberry, D.~Eriksson, A.~V.
  Evfimievski, F.~M. Manshadi, N.~Pansare, B.~Reinwald, F.~R. Reiss, and
  P.~Sen.
\newblock {SystemML: Declarative Machine Learning on Spark}.
\newblock {\em Proceedings of the VLDB Endowment}, 9(13):1425--1436, sep 2016.

\bibitem{DBLP:journals/ftml/BubeckC12}
S.~Bubeck and N.~Cesa-Bianchi.
\newblock {Regret Analysis of Stochastic and Nonstochastic Multi-armed Bandit
  Problems}.
\newblock {\em Foundations and Trends in Machine Learning}, 5(1):1--122, 2012.

\bibitem{Das2013}
S.~Das, V.~R. Narasayya, F.~Li, and M.~Syamala.
\newblock {CPU sharing techniques for performance isolation in multi-tenant
  relational database-as-a-service}.
\newblock {\em Proceedings of the VLDB Endowment}, 7(1):37--48, sep 2013.

\bibitem{DavidShue2012}
{David Shue}, {Michael J. Freedman}, and {Anees Shaikh}.
\newblock {Performance Isolation and Fairness for Multi-Tenant Cloud Storage |
  USENIX}.
\newblock In {\em OSDI}, 2012.

\bibitem{DelgadoCBA14}
M.~F. Delgado, E.~Cernadas, S.~Barro, and D.~G. Amorim.
\newblock Do we need hundreds of classifiers to solve real world classification
  problems?
\newblock {\em Journal of Machine Learning Research}, 15(1):3133--3181, 2014.

\bibitem{Desautels2014}
T.~Desautels, A.~Krause, and J.~W. Burdick.
\newblock {Parallelizing Exploration-Exploitation Tradeoffs in Gaussian Process
  Bandit Optimization}.
\newblock {\em Journal of Machine Learning Research}, 15:4053--4103, 2014.

\bibitem{Feurer2015}
M.~Feurer, A.~Klein, K.~Eggensperger, J.~Springenberg, M.~Blum, and F.~Hutter.
\newblock {Efficient and Robust Automated Machine Learning}.
\newblock In {\em NIPS}, pages 2962--2970, 2015.

\bibitem{Golovin2017}
D.~Golovin, B.~Solnik, S.~Moitra, G.~Kochanski, J.~E. Karro, and D.~Sculley.
\newblock {Google Vizier: A Service for Black-Box Optimization}.
\newblock In {\em KDD}, 2017.

\bibitem{Goovaerts1997}
P.~Goovaerts.
\newblock {\em {Geostatistics for natural resources evaluation}}.
\newblock Oxford University Press, 1997.

\bibitem{2017arXiv170602677G}
P.~{Goyal}, P.~{Doll{\'a}r}, R.~{Girshick}, P.~{Noordhuis}, L.~{Wesolowski},
  A.~{Kyrola}, A.~{Tulloch}, Y.~{Jia}, and K.~{He}.
\newblock {Accurate, Large Minibatch SGD: Training ImageNet in 1 Hour}.
\newblock {\em ArXiv e-prints}, June 2017.

\bibitem{gpyopt2016}
GPyOpt.
\newblock {{\{}GPyOpt{\}}: A Bayesian Optimization framework in python}.
\newblock $\backslash$url{\{}http://github.com/SheffieldML/GPyOpt{\}}, 2016.

\bibitem{Hellerstein:2012:VLDB}
J.~M. Hellerstein, C.~R{\'{e}}, F.~Schoppmann, D.~Z. Wang, E.~Fratkin,
  A.~Gorajek, K.~S. Ng, C.~Welton, X.~Feng, K.~Li, and A.~Kumar.
\newblock {The MADlib Analytics Library: Or MAD Skills, the SQL}.
\newblock {\em Proc. VLDB Endow.}, 2012.

\bibitem{Hutter2011}
F.~Hutter, H.~H. Hoos, and K.~Leyton-Brown.
\newblock {Sequential Model-Based Optimization for General Algorithm
  Configuration}.
\newblock In {\em LION}, pages 507--523. Springer-Verlag, 2011.

\bibitem{JonathanMace2015}
{Jonathan Mace}, {Peter Bodik}, {Rodrigo Fonseca}, and {Madanlal Musuvathi}.
\newblock {Retro: Targeted Resource Management in Multi-tenant Distributed
  Systems | USENIX}.
\newblock In {\em NSDI}, 2015.

\bibitem{Kingma2014}
D.~P. Kingma and J.~Ba.
\newblock {Adam: A Method for Stochastic Optimization}.
\newblock {\em ICLR}, dec 2014.

\bibitem{Kotthoff2017}
L.~Kotthoff, C.~Thornton, H.~H. Hoos, F.~Hutter, and K.~Leyton-Brown.
\newblock {Auto-WEKA 2.0: Automatic model selection and hyperparameter
  optimization in WEKA}.
\newblock {\em Journal of Machine Learning Research}, 18(25):1--5, 2017.

\bibitem{Krebs2014}
R.~Krebs, S.~Spinner, N.~Ahmed, and S.~Kounev.
\newblock {Resource Usage Control in Multi-tenant Applications}.
\newblock In {\em 2014 14th IEEE/ACM International Symposium on Cluster, Cloud
  and Grid Computing}, pages 122--131. IEEE, may 2014.

\bibitem{Krishnan2017}
S.~Krishnan and E.~Wu.
\newblock {PALM: Machine Learning Explanations For Iterative Debugging}.
\newblock In {\em Proceedings of the 2nd Workshop on Human-In-the-Loop Data
  Analytics - HILDA'17}, pages 1--6, New York, New York, USA, 2017. ACM Press.

\bibitem{Kushner1964}
H.~J. Kushner.
\newblock {A New Method of Locating the Maximum Point of an Arbitrary Multipeak
  Curve in the Presence of Noise}.
\newblock {\em Journal of Basic Engineering}, 86(1), 1964.

\bibitem{Low2012}
Y.~Low, D.~Bickson, J.~Gonzalez, C.~Guestrin, A.~Kyrola, and J.~M. Hellerstein.
\newblock {Distributed GraphLab}.
\newblock {\em PVLDB}, 5(8):716--727, apr 2012.

\bibitem{Luo16}
G.~Luo.
\newblock A review of automatic selection methods for machine learning
  algorithms and hyper-parameter values.
\newblock {\em NetMAHIB}, 5(1):18, 2016.

\bibitem{Meng2001}
X.~Meng, J.~Bradley, B.~Yavuz, E.~Sparks, S.~Venkataraman, D.~Liu, J.~Freeman,
  D.~Tsai, M.~Amde, S.~Owen, D.~Xin, R.~Xin, M.~J. Franklin, R.~Zadeh,
  M.~Zaharia, and A.~Talwalkar.
\newblock {\em {MLlib: machine learning in apache spark}}, volume~17.
\newblock MIT Press, 2016.

\bibitem{Narasayya2015}
V.~Narasayya, I.~Menache, M.~Singh, F.~Li, M.~Syamala, and S.~Chaudhuri.
\newblock {Sharing buffer pool memory in multi-tenant relational
  database-as-a-service}.
\newblock {\em Proceedings of the VLDB Endowment}, 8(7):726--737, feb 2015.

\bibitem{Schawinski2017}
K.~Schawinski, C.~Zhang, H.~Zhang, L.~Fowler, and G.~K. Santhanam.
\newblock {Generative Adversarial Networks recover features in astrophysical
  images of galaxies beyond the deconvolution limit}.
\newblock {\em Monthly Notices of the Royal Astronomical Society: Letters},
  120(1):slx008, jan 2017.

\bibitem{Snoek2012}
J.~Snoek, H.~Larochelle, and R.~P. Adams.
\newblock {Practical Bayesian Optimization of Machine Learning Algorithms}.
\newblock In {\em NIPS}, pages 2951--2959, 2012.

\bibitem{SnoekLA12}
J.~Snoek, H.~Larochelle, and R.~P. Adams.
\newblock Practical bayesian optimization of machine learning algorithms.
\newblock pages 2960--2968, 2012.

\bibitem{Sparks2015}
E.~R. Sparks, A.~Talwalkar, D.~Haas, M.~J. Franklin, M.~I. Jordan, and
  T.~Kraska.
\newblock {Automating model search for large scale machine learning}.
\newblock In {\em Proceedings of the Sixth ACM Symposium on Cloud Computing -
  SoCC '15}, pages 368--380, New York, New York, USA, 2015. ACM Press.

\bibitem{Sparks2017}
E.~R. Sparks, S.~Venkataraman, T.~Kaftan, M.~J. Franklin, and B.~Recht.
\newblock {KeystoneML: Optimizing Pipelines for Large-Scale Advanced
  Analytics}.
\newblock {\em ICDE}, 2017.

\bibitem{SrinivasKKS10}
N.~Srinivas, A.~Krause, S.~Kakade, and M.~W. Seeger.
\newblock Gaussian process optimization in the bandit setting: No regret and
  experimental design.
\newblock In {\em ICML}, pages 1015--1022, 2010.

\bibitem{Swersky2013}
K.~Swersky, J.~Snoek, and R.~P. Adams.
\newblock {Multi-Task Bayesian Optimization}.
\newblock In {\em NIPS}, pages 2004--2012, 2013.

\bibitem{Tamagnini2017}
P.~Tamagnini, J.~Krause, A.~Dasgupta, and E.~Bertini.
\newblock {Interpreting Black-Box Classifiers Using Instance-Level Visual
  Explanations}.
\newblock In {\em Proceedings of the 2nd Workshop on Human-In-the-Loop Data
  Analytics - HILDA'17}, pages 1--6, New York, New York, USA, 2017. ACM Press.

\bibitem{Thornton2013}
C.~Thornton, F.~Hutter, H.~H. Hoos, and K.~Leyton-Brown.
\newblock {Auto-WEKA}.
\newblock In {\em Proceedings of the 19th ACM SIGKDD international conference
  on Knowledge discovery and data mining - KDD '13}, page 847, New York, New
  York, USA, 2013. ACM Press.

\bibitem{Varma2017}
P.~Varma, D.~Iter, C.~{De Sa}, and C.~R{\'{e}}.
\newblock {Flipper: A Systematic Approach to Debugging Training Sets}.
\newblock In {\em Proceedings of the 2nd Workshop on Human-In-the-Loop Data
  Analytics - HILDA'17}, pages 1--5, New York, New York, USA, 2017. ACM Press.

\bibitem{Zhang2017a}
C.~Zhang, W.~Wu, and T.~Li.
\newblock {An Overreaction to the Broken Machine Learning Abstraction: The
  ease.ml Vision}.
\newblock In {\em Proceedings of the 2nd Workshop on Human-In-the-Loop Data
  Analytics - HILDA'17}, pages 1--6, New York, New York, USA, 2017. ACM Press.

\bibitem{zhang2016zipml}
H.~Zhang, K.~Kara, J.~Li, D.~Alistarh, J.~Liu, and C.~Zhang.
\newblock {ZipML: An End-to-end Bitwise Framework for Dense Generalized Linear
  Models}.
\newblock {\em ICML}, 2017.

\end{thebibliography}

%
%
%
%
%
%
%
%
%
%
%
%
%
%
%
%
%
%
%
%
%
%
%
%
%
%
%
%
%
%
%
%
%
%
%
%
%
%
%
%
%
%
%
%
%
%


\appendix

\section{Detailed Experiment Settings}

GP-UCB requires to specify a prior $(\mu(\mathbf{x}), k(\mathbf{x},\mathbf{x}'))$.
As a convention, for GP's not conditioned on data, we assume that $\mu=0$ without loss of generality~\cite{SrinivasKKS10}.
While we can use standard kernel functions for the $k$, we need to come up with a vector representation (i.e., the $\mathbf{x}$) for the models.
For this purpose, we randomly split a dataset into a ``training set'' and a ``testing set.''
We use the training set to compute a vector for each model as follows:
We first evaluate the model on each user in the training set to get its quality, and we then pack these qualities into a ``quality vector'' $\mathbf{x}$ indexed by the users.
In our experiments, we used 90\% data for training and 10\% data for testing.
We run each experiment 50 times by using different random splits of the data into training and testing sets.

Given a dataset, for each of the three algorithms, we report its \emph{accuracy loss} as time goes by.
For a user $i\in[n]$ and a time step $T$, let $a_{i,T}=\max_{1\leq t\leq T} a_{i,t}$ be the best accuracy of evaluating the selected models up to the time $T$, and let $a_i^{*}$ be the best accuracy that the user $i$ can achieve.
We define the accuracy loss of the user $i$ at the time $T$ as
\begin{equation}
l_{i,T}=a_i^{*}-a_{i,T}.
\end{equation}
Accordingly, we define the accuracy loss at the time $T$ to be the average accuracy loss of the users:
\begin{equation}\label{eq:accuracy-loss}
l_T=\frac{1}{n}\sum\nolimits_{i=1}^{n}l_{i,T}.
\end{equation}
Clearly $l_T\to 0$ as $T\to\infty$, if the multi-tenant model selection algorithm is regret-free (i.e., the best model for each user is eventually found).
Moreover, we have $l_{i,T}\leq r_{i,T}$, namely, the accuracy loss of each user is upper bounded by his/her instantaneous regret.

One may wonder why we report accuracy loss instead of regret.
The reason is that accuracy loss is a metric we care more in practice.
Recall the real scenario where the multi-tenant model selection algorithm is applied in our system (Section~\ref{sec:architecture}): At each time step $T$, a user is using the best model he/she has found so far (even if he/she is not scheduled at time $T$) for his/her task, rather than using the model being evaluated at time $T$!
Therefore, compared with regret, accuracy loss is closer in the sense of characterizing the real user ``happiness.''
On the other hand, regret is the standard metric when analyzing bandit algorithms.
Given that accuracy loss is upper bounded by instantaneous regret, the upper bounds for regret apply to accuracy loss as well.

\section{Details of Synthetic Data Generation}\label{sec:appendix:synthetic}

We present the model we used to generate synthetic data for $N$ users and $M$ models. The dataset we used is a special case
of this model. The basic assumption behind our model is
that the quality of a model $j\in[M]$ for a user $i\in[N]$ can be decomposed into the following four factors:

\vspace{-0.5em}
\begin{enumerate}
\item {\bf User baseline quality} $b_i$: Different users have different
      hardness associated with their tasks --- we can achieve higher accuracy on some tasks than the others.
      For a user $i$, $b_i$ then describes how difficult the corresponding task is.
\vspace{-0.5em}
\item {\bf Model quality correlation} $m_j$: For each user $i$, $\{m_j\}$ corresponds to the fluctuation of the quality of the model $j$ over the baseline quality that can be explained by model correlation.
      For example, on a given user dataset, we would expect ResNet-20 always outperforms ResNet-50 if the dataset is small.
\vspace{-0.5em}
\item {\bf User quality correlation} $u_i$: For each model $j$, $\{u_i\}$ corresponds to the fluctuation of the quality of the model $j$ across different users
      over the baseline quality that can be explained by user dataset correlation.
      For example, given a neural network such as ResNet-1000, we should expect that it works well on all large datasets but worse on all small ones.
\vspace{-0.5em}
\item {\bf White noise} $\epsilon_{i,j}$: This factor tries to capture everything that are
      not explained by the previous three factors.
\end{enumerate}
\vspace{-0.5em}

Let $x_{i,j}$ be the quality of model $j$ for user $i$. With the above notation, it follows that
\begin{equation}\label{eq:quality}
x_{i,j} = b_i + m_j + u_i + \epsilon_{i,j}.
\end{equation}
By convention we define $x_{i,j}=0$ if Equation~\ref{eq:quality} leads to $x_{i,j}<0$ and define $x_{i,j}=1$ if Equation~\ref{eq:quality} leads to $x_{i,j}>1$.

The generative procedure for each $x_{i,j}$ is to draw samples 
$b_i$, $m_j$, $u_i$, $\epsilon_{i,j}$ from their corresponding 
distributions (defined as follows), and then sum them together.

\subsection{Generative Model}

The generative model behind the synthetic dataset we used is as follows.

\begin{enumerate}
\item {\bf Baseline Group:} There are different groups of tasks -- those that are difficult (with small
baseline quality) and those that are easy (with higher baseline quality). A given user can belong
to a single baseline group.
\item {\bf User Group:} A user group is a group of users who share the
same generative model of $u_i$. There are different groups -- high correlation groups
and low correlation groups. A given user can belong to a single user group.
\item {\bf Model Group:} A model group is a group of models who share
the same generative model of $m_j$. There are different groups -- high correlation groups
and low correlation groups. A given model can belong to a single model group.
\item {\bf White Noise:} We assume all white noises are i.i.d.
\end{enumerate}

Given a user who belongs to baseline group $B$ and user group $U$,
and a model that belongs to model group $M$, each of the above four terms
can be sampled from the corresponding distribution of the group.

\subsubsection{Generative Model of a Baseline Group}

Different baseline groups are governed by the expected quality $\mu_b$ of the group
and the variation of quality $\sigma_b$ within the group. Given a user
belong to a $(\mu_b, \sigma_b)$-baseline group, the user baseline quality 
$b$ is sampled as
\[
b \sim \mathcal{N}(\mu_b, \sigma_b).
\]

\subsubsection{Generative Model of a Model Group}

Different model groups are governed by a constant $\sigma_M$ corresponding to
the strength of correlation. Let ${M_1,...,M_K}$ be all models in a
$\sigma_M$-model group, we assume the following generative models.

\paragraph*{Hidden Similarity between Models}

One assumption we have is that each model has a similarity measure with each
other that is {\em not known to the algorithm}. This similarity defines how strong
the correlations are between different models. For each 
model $M_j$, we assign 
\[
f(M_j) \sim \mathcal{\mathcal{U}}(0, 1)
\]
and will use it later to calculate the similarity between models.

\paragraph*{Sample from a model group}

Given a $\sigma_M$-model group and the $f(-)$ score already pre-assigned to each
model, we sample 
\[
[m_1,...m_K] \sim \mathcal{N}(0, \Sigma_M)
\]
where the covariance matrix $\Sigma_M$ is defined as follows.

For each pair of $(i,j)$, $\Sigma_M[i,j]$ follows the following intuition:
If $f(M_i)$ and $f(M_j)$ are close to each other, then their correlation
is stronger. Thus, we define
\[
\Sigma_M[i,j] = \exp \left\{-\frac{(f(M_i) - f(M_j))^2}{\sigma_M^2} \right\}
\]

\subsubsection{Generative Model of the User Group}

Different user groups are governed by a constant $\sigma_U$ corresponding to
the strength of correlation. Let ${M_1,...,M_K}$ be all users in a
$\lambda_U$-user group, the generative model is similar to that of a model group.
\[
[u_1,...,u_K] \sim \mathcal{N}(0, \Sigma_U).
\]

\subsubsection{Generative Model of the White Noise}

All white noises are i.i.d, governed by the same constant $\sigma_W$.
\[
\epsilon_{i,j} \sim \mathcal{N}(0, \sigma_W).
\]

\subsection{Dataset Generation}

A synthetic dataset is governed by the following tuple
\begin{eqnarray*}
( 
\mathcal{B} = \{(\mu_b^{(i)}, \sigma_b^{(i)})\},
\mathcal{M} = \{(\sigma_M^{(j)})\},
\mathcal{U} = \Phi,
\sigma_W,  \\
p_U: \mathcal{B} \times \mathcal{U} \mapsto \mathbb{N},
p_M: \mathcal{M} \mapsto \mathbb{N}
)
\end{eqnarray*}
which is generated by 
$|\mathcal{B}|$ baseline groups, 
$|\mathcal{M}|$ model groups, 
and $|\mathcal{U}|$ user groups. $p_U$
maps each combination of baseline group
and user group to the number of users
belonging to the given combination.
$p_M$ maps each model group to how many
models belong to the given group.

The synthetic dataset we used
in this paper uses the following instantiation.

\begin{enumerate}
\item $\mathcal{B} = \{(0.75, \sigma_{B}), (0.25, \sigma_{B})\}$
\item $\mathcal{M} = \{\sigma_{M}\}$
\item $\mathcal{U} = \{\sigma_U\}$
\item $p_U(*)$ = 50
\item $p_M(*)$ = 100
\end{enumerate}
which will generate 100 models and 100 users. 

To understand the impact of different factors to the algorithm,
\begin{enumerate}
\item Change $\sigma_{M}$ to understand the impact of
correlations between models.
\item Change $\sigma_{U}$ to understand the impact of
correlations between users.
\item Change $P_U$ to understand the impact of
mixing users with different task difficulties (if all users' task are
equally hard, why not round-robin)
\end{enumerate}

\section{Proofs}
In the sequel, we denote by $f^i(k)$ the performance or reward of choosing model $k\in [K]$ for tenant $i\in [n]$, and for single tenant case, we simply use $f(k)$.
\begin{lemma}
  \label{lem:5.1}
  Fix $\delta \in (0,1)$ and let $\beta_t = 2c^\ast \log\left[\frac{\pi^2 K t^2}{6\delta}\right]$. Then
  \[
    |f(k) - \mu_{t-1}(k) | \le \sqrt{\frac{\beta_t}{c_k}}\sigma_{t-1}(k),\
    \forall 2\le t\le T, k\in [K],
  \]
  with probability at least $1-\delta$.
\end{lemma}

\begin{proof}
  Fix $t\ge 2$. Conditioned on $y_{[1:t-1]}$, we have
  \[
    f(k) \sim \cN(\mu_{t-1}(k), \sigma^2_{t-1}(k)), \ \forall k\in [K].
  \]
  Then
  \begin{align*}
    & \bP\left(|f(k)- \mu_{t-1}(k)| \ge \sqrt{\frac{\beta_t}{c_k}}\sigma_{t-1}(k)\right)\\
    =\ & \bP\left(|f(k)- \mu_{t-1}(k)|\sigma^{-1}_{t-1}(k) \ge \sqrt{\frac{\beta_t}{c_k}}\right)\\
    =\ & \frac{2}{\sqrt{2\pi}}\int_{\sqrt{\frac{\beta_t}{c_k}}}^\infty e^{-\frac{z^2}{2}}\, dz = e^{-\frac{\beta_t}{2c_k}} \frac{1}{\sqrt{2\pi}}\int_{\sqrt{\frac{\beta_t}{c_k}}}^\infty e^{-\frac{\beta_t/c_k-z^2}{2}}\, dz\\
    =\ & e^{-\frac{\beta_t}{2c_k}} \frac{2}{\sqrt{2\pi}}\int_{\sqrt{\frac{\beta_t}{c_k}}}^\infty e^{-\frac{(\sqrt{\beta_t/c_k}-z)^2}{2}}e^{-z(z-\sqrt{\beta_t/c_k})}\, dz\\
    \le \ & e^{-\frac{\beta_t}{2c_k}} \frac{2}{\sqrt{2\pi}} \int_0^\infty e^{-\frac{z^2}{2}}\, dz\\
    =\ & e^{-\frac{\beta_t}{2c_k}}.
  \end{align*}
  Applying the union bound for $2\le t\le T, k\in [K]$, we obtain
  \[
    |f(k) - \mu_{t-1}(k) | \le \sqrt{\frac{\beta_t}{c_k}}\sigma_{t-1}(k),\
    \forall k\in [K],
  \]
  with probability at least $1- \sum_{t=1}^T\sum_{k=1}^K
  e^{-\frac{\beta_t}{2c_k}} \ge  1 - \delta$.
\end{proof}

\vspace{2em}
\paragraph*{Proof of Theorem \ref{thm:cost}}

\begin{proof}
We here only show the proof of the bound for the simple regret, and the cumulative regret case can be shown by setting $n=1$ in the proof of multi-tenant case.

By the definition of $a_t$ in our algorithm, we have
  \[
    \mu_{t-1}(a_t) + \sqrt{\frac{\beta_t}{c_{a_t}}}\sigma_{t-1}(a_t) \ge
    \mu_{t-1}(a^\ast) + \sqrt{\frac{\beta_t}{c_{a^\ast}}}\sigma_{t-1}(a^\ast).
  \]
  Then it follows from Lemma \ref{lem:5.1} that with at least probability
  $1-\delta$, we have
  \begin{align*}
    r_t & = f(a^\ast) - f(a_t)\\
        & \le \mu_{t-1}(a^\ast) + \sqrt{\frac{\beta_t}{c_{a^\ast}}}\sigma_{t-1}(a^\ast) - f(a_t)\\
        & \le \mu_{t-1}(a_t) + \sqrt{\frac{\beta_t}{c_{a_t}}}\sigma_{t-1}(a_t) - f(a_t)\\
    & \le 2 \sqrt{\frac{\beta_t}{c_{a_t}}}\sigma_{t-1}(a_t).
  \end{align*}
  Note that $\beta_t$ is increasing in $t$, $\sigma_{t-1}^2(a_t) \le \Sigma(a_t,a_t)\le 1$ and the function
  $x/\log(1+x)$ is increasing in $x\ge0$, we have
  \begin{align*}
    c_{a_t} r_t^2 &\le c_{a_t} \cdot 4\beta_t c^{-1}_{a_t} \sigma_{t-1}^2(a_t) = 4\beta_t \sigma_{t-1}^2(a_t)\\
                  & \le 4\sigma^2 \beta_T \frac{\sigma^{-2}\sigma^2_{t-1}(a_t)}{\log(1+\sigma^{-2}\sigma^2_{t-1}(a_t))}\cdot \log(1+\sigma^{-2}\sigma^2_{t-1}(a_t))\\
                  & \le 4\sigma^2 \beta_T \frac{\sigma^{-2}}{\log(1 + \sigma^{-2})}\cdot \log(1+\sigma^{-2}\sigma^2_{t-1}(a_t))\\
                  & = \frac{4\beta_T}{\log(1 + \sigma^{-2})}\cdot \log(1+\sigma^{-2}\sigma^2_{t-1}(a_t)),
  \end{align*}
  and thus
  \[
    \sum_{t=1}^Tc_{a_t} r_t^2 \le \tilde{I}(T).
  \]
  By Cauchy-Schwarz inequality, we obtain
  \begin{align*}
    \min_{t\in [T]}r_t & 
                         \le \frac{\sum_{t=1}^Tc_{a_t}r_t}{\sum_{t=1}^Tc_{a_t}}\le \frac{\sqrt{\sum_{t=1}^Tc_{a_t}}\cdot \sqrt{\sum_{t=1}^Tc_{a_t} r_t^2}}{\sum_{t=1}^Tc_{a_t}}\\
& \le \sqrt{\frac{\tilde{I}(T)}{\sum_{t=1}^T c_{a_t}}},
  \end{align*}
which completes the proof.
\end{proof}

\vspace{2em}
\paragraph*{Proof of Theorem \ref{thm:rr}}
\begin{proof}
   Fix $t\ge 2$ and $i\in [n]$. Conditioned on the observations of previous $t-1$ rounds, we
  know that 
  \[
    f^i(k) \sim \cN(\mu^i_{t-1}(k), (\sigma^i_{t-1}(k))^2), \ \forall k\in [K^i].
  \]
  Then
  \begin{align*}
    & \bP\left(|f^i(k)- \mu^i_{t-1}(k)| \ge \sqrt{\frac{\beta^i_t}{c^i_k}}\sigma^i_{t-1}(k)\right)\\
    =\ & \bP\left(\frac{|f^i(k)- \mu^i_{t-1}(k)|}{\sigma^i_{t-1}(k)} \ge \sqrt{\frac{\beta^i_t}{c^i_k}}\right)\\
     \le\ & e^{-\frac{\beta^i_t}{2c^i_k}}.
  \end{align*}
  Applying the union bound, we obtain for any $2\le t\le T, i\in [n]$ and $k\in [K^i]$,
  \[
    |f^i(k) - \mu^i_{t-1}(k) | \le \sqrt{\frac{\beta^i_t}{c^i_k}}\sigma^i_{t-1}(k),
  \]
  with probability at least $1- \sum_{t=1}^T\sum_{i=1}^n\sum_{k=1}^{K^i}
  e^{-\frac{\beta^i_t}{2c^i_k}} \ge  1 - \delta$.

  For each user $i\in [n]$, we have
  \begin{align*}
    r^i_{t^i} & = f^i(a^{i,\ast}) - f^i(a^i_{t^i})\\
    & \le \sqrt{\frac{\beta^i_{t^i}}{c^i_{a^i_{t}}}}\sigma^i_{t-1}(a^i_{t^i}) +\mu^i_{t-1}(a^i_{t^i}) - f^i(a^i_{t^i})\\
    & \le 2
    \sqrt{\frac{\beta^i_{t^i}}{c^i_{a^i_{t}}}}\sigma^i_{t-1}(a^i_{t^i}),
  \end{align*}
   and thus by algorithm,
  \begin{align*}
    \sum_{t=1}^T \sum_{i=1}^n c^{I_t}_{a^{I_t}_t} r^i_{t^i}
    & \le \sum_{t=1}                                                    ^T \sum_{i=1}^n c^{I_t}_{a^{I_t}_t} 2 \sqrt{\frac{\beta^i_{t^i}}{c^i_{a^i_{t^i}}}}\sigma^i_{t-1}(a^i_{t^i})\\
        & \le 2 \sqrt{\frac{(c^\ast)^2}{c_\ast}} \sum_{t=1}^T \sum_{i=1}^n \sqrt{\beta^i_{t^i}}\sigma^i_{t-1}(a^i_{t^i})\\
    & = 2 n \sqrt{\frac{(c^\ast)^2\beta^\ast}{c_\ast}}\sum_{i=1}^n\sum_{t\in T(i)} \sigma^i_{t-1}(a^i_{t})\\
    & \le 2 n \sqrt{\frac{(c^\ast)^2\beta^\ast}{c_\ast}}\sum_{i=1}^n\sqrt{|T(i)|}\sqrt{\sum_{t\in T(i)} (\sigma^i_{t-1}(a^i_{t}))^2}\\
    & \le 2 n \sqrt{\frac{(c^\ast)^2\beta^\ast}{c_\ast}}\sum_{i=1}^n\sqrt{\frac{2T}{n}}\sqrt{\sum_{t\in T(i)} (\sigma^i_{t-1}(a^i_{t}))^2}
  \end{align*}
  Note that
\begin{align*}
    (\sigma^i_{t-1}(a^i_{t}))^2
    & = (\sigma^i)^2\cdot \frac{(\sigma^i)^{-2} (\sigma^i_{t-1}(a^i_{t}))^2}{\log\left(1 + (\sigma^i)^{-2} (\sigma^i_{t-1}(a^i_{t}))^2\right)}\\
    &\qquad \cdot \log\left(1 + (\sigma^i)^{-2} (\sigma^i_{t-1}(a^i_{t}))^2\right)\\
    & \le (\sigma^i)^2 \cdot \frac{(\sigma^i)^{-2}}{\log\left(1 + (\sigma^i)^{-2}\right)}\\
    &\qquad \cdot \log\left(1 + (\sigma^i)^{-2} (\sigma^i_{t-1}(a^i_{t}))^2\right)\\
    & = \frac{1}{\log\left(1 + (\sigma^i)^{-2}\right)}\\   
    &\qquad \cdot \log\left(1 + (\sigma^i)^{-2} (\sigma^i_{t-1}(a^i_{t}))^2\right),
  \end{align*}
  and so
  \begin{align*}
      \sum_{t\in T(i)} (\sigma^i_{t-1}(a^i_{t}))^2
      & \le \frac{1}{\log\left(1 + (\sigma^i)^{-2}\right)}\\
    &\qquad \sum_{t\in T(i)} \log\left(1 + (\sigma^i)^{-2} (\sigma^i_{t-1}(a^i_{t}))^2\right).
  \end{align*}
  Therefore,
  \[
    \sum_{t=1}^T \sum_{i=1}^n c^{I_t}_{a^{I_t}_t} r^i_{t^i} \le \sqrt{n T}\sum_{i=1}^n \sqrt{I_i([T(i)])}.
  \]
\end{proof}

\vspace{2em}
\paragraph*{Proof of Theorem \ref{thm:multi-cost}}
\begin{proof}
  Fix $t\ge 2$ and $i\in [n]$. Conditioned on the observations of previous $t-1$ rounds, we
  know that 
  \[
    f^i(k) \sim \cN(\mu^i_{t-1}(k), (\sigma^i_{t-1}(k))^2), \ \forall k\in [K^i].
  \]
  Then
  \begin{align*}
    & \bP\left(|f^i(k)- \mu^i_{t-1}(k)| \ge \sqrt{\frac{\beta^i_t}{c^i_k}}\sigma^i_{t-1}(k)\right)\\
    =\ & \bP\left(\frac{|f^i(k)- \mu^i_{t-1}(k)|}{\sigma^i_{t-1}(k)} \ge \sqrt{\frac{\beta^i_t}{c^i_k}}\right)\\
    \le\ & e^{-\frac{\beta^i_t}{2c^i_k}}.
  \end{align*}
  Applying the union bound, we obtain for any $2\le t\le T, i\in [n]$ and $k\in [K^i]$,
  \[
    |f^i(k) - \mu^i_{t-1}(k) | \le \sqrt{\frac{\beta_t^i}{c^i_k}}\sigma^i_{t-1}(k),
  \]
  with probability at least $1- \sum_{t=1}^T\sum_{i=1}^n\sum_{k=1}^{K^i}
  e^{-\frac{\beta^i_t}{2c^i_k}} \ge  1 - \delta$.
 
  It follows from the algorithm, we have
  \begin{align*}
    r^i_{t^i} & = f^i(a^{i,\ast}) - f^i(a^i_{t_i})\\
    & \le \tilde{\sigma}^i_{t^i} \le 2
    \sqrt{\frac{\beta^i_{t^i}}{c^i_{a^i_{t^i}}}}\sigma^i_{t-1}(a^i_{t^i}),\ \forall
    i\in [n], t\in [T],
  \end{align*}
  and thus,
  \begin{align*}
    \sum_{t=1}^T \sum_{i=1}^n c^{I_t}_{a^{I_t}_t} r^i_{t^i}
    & \le \sum_{t=1}^T c^{I_t}_{a^{I_t}_t} \sum_{i=1}^n  2 \sqrt{\frac{\beta^i_{t^i}}{c^i_{a^i_{t^i}}}}\sigma^i_{t-1}(a^i_{t^i})\\
        & \le 2n  \sum_{t=1}^T c^{I_t}_{a^{I_t}_t} \sqrt{\frac{\beta^{I_t}_{t}}{c^{I_t}_{a^{I_t}_{t}}}} \sigma^{I_t}_{t-1}(a^{I_t}_t) \quad(\text{since}~t^{I_t}=t)\\
    & = 2 n \sum_{t=1}^T \sqrt{c^{I_t}_{a^{I_t}_{t}}\beta^{I_t}_{t}} \sigma^{I_t}_{t-1}(a^{I_t}_t) \\
    & \le 2n \sqrt{\sum_{t=1}^Tc^{I_t}_{a^{I_t}_{t}}\beta^{I_t}_{t}}\sqrt{\sum_{t=1}^T (\sigma^{I_t}_{t-1}(a^{I_t}_t))^2}\\
    & \le 2n \sqrt{c^\ast\beta^\ast T}\sqrt{\sum_{t=1}^T (\sigma^{I_t}_{t-1}(a^{I_t}_t))^2}.
  \end{align*}
  Note that
  \begin{align*}
    \sum_{t=1}^T (\sigma^{I_t}_{t-1}(a^{I_t}_t))^2
    & = \sum_{i=1}^n \sum_{t\in T(i)} (\sigma^i_{t-1}(a^i_{t}))^2,
  \end{align*}
  and also
  \begin{align*}
    (\sigma^i_{t-1}(a^i_{t}))^2
    & = (\sigma^i)^2\cdot \frac{(\sigma^i)^{-2} (\sigma^i_{t-1}(a^i_{t}))^2}{\log\left(1 + (\sigma^i)^{-2} (\sigma^i_{t-1}(a^i_{t}))^2\right)}\\
    &\qquad \cdot \log\left(1 + (\sigma^i)^{-2} (\sigma^i_{t-1}(a^i_{t}))^2\right)\\
    & \le (\sigma^i)^2 \cdot \frac{(\sigma^i)^{-2}}{\log\left(1 + (\sigma^i)^{-2}\right)}\\
    &\qquad \cdot \log\left(1 + (\sigma^i)^{-2} (\sigma^i_{t-1}(a^i_{t}))^2\right)\\
    & = \frac{1}{\log\left(1 + (\sigma^i)^{-2}\right)}\\   
    &\qquad \cdot \log\left(1 + (\sigma^i)^{-2} (\sigma^i_{t-1}(a^i_{t}))^2\right).
  \end{align*}
  Therefore,
  \begin{align*}
    \sum_{t=1}^T (\sigma^{I_t}_{t-1}(a^{I_t}_t))^2
    & \le \sum_{i=1}^n \frac{1}{\log\left(1 + (\sigma^i)^{-2}\right)}\\
    &\qquad \sum_{t\in T(i)} \log\left(1 + (\sigma^i)^{-2} (\sigma^i_{t-1}(a^i_{t}))^2\right).
  \end{align*}
  Finally, we obtain the bound for the total regret:
  \begin{align*}
    & \sum_{t=1}^T \sum_{i=1}^n c^{I_t}_{a^{I_t}_t} r^i_{t^i}\\
     \le \ & 2n \sqrt{c^\ast \beta^\ast T}\\
    & \cdot \sqrt{\sum_{i=1}^n \frac{1}{\log\left(1 + (\sigma^i)^{-2}\right)} \sum_{t\in T(i)} \log\left(1 + (\sigma^i)^{-2} (\sigma^i_{t-1}(a^i_{t}))^2\right)}\\
    =\ & n \sqrt{T} \sqrt{\sum_{i=1}^n I_i([T(i)])},
  \end{align*}
 which completes the proof.
\end{proof}


\end{document}